\documentclass[useAMS,usenatbib]{mn2e}
\usepackage{deluxetable}
\usepackage{epsf}
\usepackage{amsmath}

\newcommand{\kmsmpc}{\kms\;{\rm Mpc}^{-1}}

\newcommand{\hkpc}{h^{-1}{\rm kpc}}
\newcommand{\hmpc}{h^{-1}{\rm Mpc}}

\newcommand{\kms}{\;{\rm km}\,{\rm s}^{-1}}

\newcommand{\msun}{M_{\odot}}
\newcommand{\mstar}{M_*}
\newcommand{\sfr}{\dot{M}_{\rm{SFR}}}
\newcommand{\acc}{\dot{M}_{\rm{ACC}}}
\newcommand{\mlf}{\eta_{\rm{W}}}
\newcommand{\vw}{V_{\rm{W}}}
\newcommand{\emlf}{\tilde{\eta}_{\rm{MLF}}}
\newcommand{\zeq}{Z_{g,\rm{eq}}}
\newcommand{\zsuv}{Z_{*,\rm{UV}}}
\newcommand{\zsall}{Z_{*,\rm{all}}}
\newcommand{\fzret}{f_{Z,\rm{ret}}}

\begin{document}

\title{The Origin of the Galaxy Mass-Metallicity Relation and Implications for Galactic Outflows}

\author[Finlator \& Dav\'e]{Kristian Finlator \& Romeel Dav\'e\\Astronomy Department, University of Arizona, Tucson, AZ 85721}

\maketitle

 \begin{abstract}
Using cosmological hydrodynamic simulations that dynamically incorporate
enriched galactic outflows together with analytical modeling, we study
the origin of the stellar mass--gas-phase metallicity relation (MZR).  
We find that metallicities are driven by an equilibrium between the 
rate of enrichment owing to star formation and the rate of dilution 
owing to infall of unenriched gas.  This equilibrium is in turn governed 
by the outflow strength.  As such, the MZR provides valuable insights and 
strong constraints on galactic outflow properties across cosmic time.  
We compare three outflow models: No outflows, a ``constant wind" 
model that emulates the popular \citet{dek86} scenario,
and a ``momentum-driven wind" model that best reproduces $z\ga 2$
intergalactic medium metallicity data~\citep{opp06}.  Only the 
momentum-driven wind scaling simulation is able to reproduce the
observed $z\sim 2$ MZR's slope, amplitude, and scatter.  In order to
understand why, we construct a one-zone chemical evolution model guided
by simulations.  This model shows that the MZR in our outflow simulations
can be understood in terms of three parameters: (1) The {\it equilibrium
metallicity} $\zeq=y\sfr/\acc$ (where $y$=net yield), reflecting the
enrichment balance between star formation rate $\sfr$ and gas accretion
rate $\acc$; (2) the {\it dilution time} $t_d=M_g/\acc$, representing the
timescale for a galaxy to return to $\zeq$ after a metallicity-perturbing
interaction; and (3) the {\it blowout mass} $M_{\rm blowout}$, which is the
galaxy stellar mass above which winds can escape its halo.  Without
outflows, galaxy metallicities exceed observations by $\sim\times 2-3$,
although the slope of the MZR is roughly correct owing to greater star
formation efficiencies in larger galaxies.  When outflows with mass
loading factor $\mlf$ are present, galaxies below $M_{\rm blowout}$
obey $\zeq\approx y/(1+\mlf)$, while above $M_{\rm blowout}$, 
$\zeq\rightarrow y$.  Our constant wind model has $M_{\rm blowout}\sim
10^{10}M_\odot$, which yields a sharp upturn in the MZR above this scale
and a flat MZR with large scatter below it, in strong disagreement with
observations.  Our momentum-driven wind model naturally reproduces the
observed $Z_g\propto M_*^{0.3}$ because $\zeq\propto\mlf^{-1}\propto
M_*^{1/3}$ when $\mlf\gg 1$ (i.e. at low masses).  The flattening of the
MZR at $M_*\ga 10^{10.5}M_\odot$ observed by~\citet{tre04} is reflective
of the mass scale where $\mlf\sim 1$, rather than a characteristic outflow
speed; in fact, the outflow speed plays little role in the MZR except
through $M_{\rm blowout}$.  The tight observed MZR scatter is ensured
when $t_d\la$~dynamical time, which is only satisified at all masses in
our momentum-driven wind model.  We also discuss secondary effects on
the MZR, such as baryonic stripping from neighboring galaxies' outflows.
\end{abstract}

 
\section{Introduction} \label{sec:intro}
In the reigning hierarchical model of galaxy formation, cooling times
for moderate overdensity gas at high redshifts ($z>3$) are shorter than a 
Hubble time, and protogalaxies accrete gas from the intergalactic medium 
(IGM) at rates that increase with time~\citep[e.g.,][]{bir07}.  Gas forms 
into stars at rates that mainly track the rising gas accretion 
rates~\citep{ker05}; these rates also scale with galaxy mass and are 
occasionally boosted by mergers~\citep[e.g.][]{som01,str06}.  Each 
generation of stars in turn enriches the surrounding gas with heavy 
elements.  The young galaxies also drive much of their enriched 
gas out via galactic superwinds, lacing the IGM with heavy 
elements~\citep{ade05b,ade03,sch03} while suppressing their own star 
formation~\citep{spr03b,opp06}.  As the universe expands further, cooling 
times rise, until $z\sim 1-2$ when the cooling time at moderate 
overdensities exceeds the Hubble time, and gas accretion rates and star 
formation rates begin to decline.

At any given epoch the accumulated history of star formation, inflows,
and outflows affects a galaxy's mass and its metallicity.  Hence one
expects these quantities (or their proxies) to be correlated in some
way, and furthermore for this correlation to encode information about
the physical processes that govern galaxy formation.  In this paper, we
investigate what constraints may be placed on such processes, particularly
outflow processes that are currently the most poorly understood, based
on the observed mass-metallicity relation (MZR) of galaxies.

\citet{mcc68} were the first to observe a correlation between the
luminosity and metallicity of elliptical galaxies, and~\citet{leq79} were
the first to show that total mass (or, equivalently, rotational velocity)
correlates with metallicity for irregular and blue compact galaxies.
The question of which relationship is more fundamental~\citep{zar94,gar97}
was resolved when~\citet{tre04} showed, for a sample of $\approx53,000$
star-forming galaxies from the Sloan Digital Sky Survey, that the MZR
possesses much less intrinsic scatter than the luminosity-metallicity
relation.  The trends seen by~\citet{tre04} have recently been shown
to extend unbroken to much lower masses by~\citet{lee06}, confirming
the idea~\citep{gar02} that a single mechanism may govern galaxies'
metallicities across five decades in stellar mass.  Finally, the MZR is
observed to evolve slowly such that galaxies of a given stellar mass
are only moderately more enriched today as compared to in the early
Universe~\citep{sha05,sav05,erb06,ber06}.

A number of mechanisms have been proposed to explain the observed trends.
Building upon an idea originally introduced by~\citet{mat71} 
and~\citet{lar74},~\citet{dek86} and~\citet{dek03} showed that supernova
feedback energy could give rise to a range of observed trends in low
mass galaxies including the MZR, if the supernova energy 
injected into galaxies' interstellar media is proportional to its
stellar mass (as one might na\"ively expect).  In this model, there is
a characteristic halo circular velocity of $\sim 100$~km~s$^{-1}$
above which galaxies retain their gas, and below which gas removal
becomes progressively more efficient.

More recent investigations have attempted to put the MZR in a hierarchical
context.  \citet{luc04} used semianalytical models to suggest that such
winds can be tuned to reproduce the $z\approx 0$ MZR irrespective of what
the outflows do upon leaving the galaxy.  \citet{ros06} and~\citet{tas06}
used cosmological hydrodynamic simulations without strong supernova 
feedback to obtain rough agreement with the~\citet{tre04} 
and~\citet{dek03} MZRs, respectively.  Both works cite the possible role of
a varying star formation efficiency with stellar mass, while~\citet{tas06}
also suggest that the observed low effective yields in low-mass galaxies
could result from mixing processes that transport metals to galaxies' outer 
disks, where they are difficult to observe.  \citet{bro06} used cosmological
hydrodynamic simulations with a treatment for pressure-driven outflows to 
argue that mass loss does not directly suppress the metallicities of low 
mass galaxies, and instead argued that supernova feedback leads to low star 
formation efficiencies in low-mass galaxies, which in turn lead to low 
metallicities; it is particularly encouraging that their model reproduces 
the observed trends both at $z\sim2$ and at $z\sim0$.  \citet{kob07} used
a similar model that also incorporated a treatment for hypernova feedback and
obtained qualitative agreement with the same observations.  However, they
concluded that their MZR is primarily driven by a tendency for low-mass 
galaxies to eject relatively more material in outflows, an idea that our 
results support.  A more speculative idea from~\citet{kop06} suggests that 
the observed trends at low redshift can be reproduced by postulating that
the stellar initial mass function (IMF) is more top-heavy in galaxies
with higher star formation rates, thereby producing higher metal yields.

In short, the galaxy MZR has been speculated as arising from variation
in mass loss, star formation efficiency, and/or yield with galaxy stellar
mass.  A key aim of this present work is to distinguish between these
alternatives, and determine the key driver that sets the MZR.  

A feature that has garnered much attention is the apparent flattening
of the MZR for $M_*\ga 10^{10.5} \msun$.  This characteristic mass
also seems to divide galaxy properties in general, such as blue from
red and high surface brightness from low~\citep{kau04}.  \citet{dek03}
note that their predicted characteristic halo velocity~\citep{dek86}
is in broad agreement with this mass.  Building on this, \citet{tre04}
and~\citet{gar02} proposed that winds may not be effective at driving
metals out of galaxies above this mass range (see however~\citealt{dal06}).
One physical model that could give rise to this behavior is the
``constant wind" scenario.  In this model, the low effective yields 
observed in small galaxies result from winds that have roughly constant 
velocities at all masses, and hence are progressively more effective at 
driving out metals from the shallower potential wells of smaller 
galaxies.  Consequently, the escape velocity at the observed MZR turnover 
should be an indicator of the characteristic wind speed.  At present, 
this scenario of mass loss variation is probably the most widely accepted 
explanation for the MZR's shape.

In this work we show that such a constant wind scenario, when incorporated
into a fully three-dimensional hierarchical structure formation
simulation, produces an MZR shape that is in poor agreement with
observations, for reasons that can be understood from straightforward
physical arguments.  This scenario has been also called into question
recently by observational analyses of dwarf galaxy metallicities.
\citet{lee06} note that such energetic speeds from small galaxies
should produce a much greater scatter in metallicities than observed in
their sample of dwarf irregulars.  They instead propose that ``a less
energetic form of metal-enhanced mass loss than blowouts could explain
the small scatter."  \citet{dal06} used a rigorous treatment of the
effects of outflows to show that outflows alone cannot account for
the low observed effective yields of dwarf galaxies, unless the winds are
substantially enriched relative to ISM gas.  She proposed instead that
the low gas surface densities of galaxies with circular velocities below
$120$ km s$^{-1}$ lead to low star formation rates (SFRs) so that their
effective yields ``recover" from outflows relatively slowly.  Hence she
suggests star formation efficiency variations are the key driver of
the MZR.  The results we present here are generally in agreement with
this conclusion.

In this paper we use cosmological hydrodynamic simulations and simple
analytical models to investigate the origin of the mass-metallicity
relation.  Our simulations employ parameterized outflows from
star-forming galaxies that drive metals into the IGM, and hence we
directly track the growth of galaxy metallicity along with its mass
within a full three-dimensional hierarchical structure formation
scenario.  In \citet{opp06} we introduced our outflow models, and
showed in particular that models in which the outflow velocity scaled
linearly with galaxy circular velocity while the mass loading factor
(i.e. the rate of mass ejection relative to the star formation rate)
scaled inversely with it were remarkably successful at enriching the IGM
to observed levels at $z\sim 2-6$.  These scalings arise naturally for
radiation or momentum-driven winds~\citep[e.g.][]{mur05}, though for
our purposes the important aspect is the scaling relations themselves
and not the physical mechanism responsible.  Interestingly, recent
observations of local galactic outflows indicate momentum-driven wind
scalings~\citep{mar05,rup05}, providing an intriguing connection between
rare local outflows and the more ubiquitous and generally stronger
outflows at $z\ga 2$.

In our preliminary study of the MZR~\citep{dav06b}, we compared various
outflow models with the $z\sim2$ MZR seen by~\citet{erb06} and examined
its evolution from $z=6\rightarrow 2$.  We found that the ``constant
wind" model as implemented by~\citet{spr03b} leads to poor agreement
with observations at $z\sim2$, while our momentum-driven wind scalings
naturally reproduce the slope and amplitude of the observed relation.
While this work provided independent support for the outflow model
concurrently favored by IGM metallicity data, it did little to address the
fundamental question of what physical processes govern the MZR's slope,
amplitude, scatter, and evolution.

In order to improve our understanding of the relationship between outflows
and the MZR, we follow a trajectory that encompasses three basic goals:\\
(1) \emph{Show that our numerical model reproduces observations at $z\sim2$.}
In this step we describe our simulations (Section~\ref{sec:sims}) and 
discuss the mass scales that our outflow models introduce.  Next we 
compare the simulated MZRs assuming different outflow models with the
observed MZR at $z=2$ (Section~\ref{sec:mmr_z2}). Our finding that
momentum-driven outflows produce the best agreement with observations
motivates a more detailed investigation into the origin of the MZR
within our simulations.\\
(2) \emph{``Boil down" our numerical model to a set of key processes and 
combine them in an analytically tractable way.} We begin this step by 
introducing a simple model that captures the main processes that impact 
the growth of galaxies' stellar masses and metallicities 
(Section~\ref{sec:analytics}).  Next, we investigate how gas accretion 
and star formation rates vary with mass and time in each outflow scenario 
in order to treat them accurately in our analytical model.  We continue 
this discussion in 
Section~\ref{sec:sfefficiency} with a detailed investigation into the 
time-integrated effects of outflows on our simulated galaxies at $z=2$.
In Section~\ref{sec:metEvol} we trace the observable trends at $z=2$ back
in time in order to understand how galaxies evolve through the MZR.  In
Section~\ref{ssec:compare_evol} we verify that our analytical model reproduces 
this evolution reasonably well, which suggests that our analytical model
accounts for the important processes that drive the observable MZR.\\
(3) \emph{Use the analytical model to determine what drives the observable MZR 
in the presence of outflows.} In Section~\ref{sec:compare_analytics} we use 
our analytical model to determine what drives the form of the observable MZR; 
the reader may wish to skip directly to this Section for a relatively 
self-contained explanation of the origin of the MZR.  Here we show why our 
momentum-driven wind model is successful at reproducing the $z\approx 2$ MZR, 
and why other models are not.  More generally, we show how the observed MZR's 
slope, amplitude, and scatter down to low masses provide stringent constraints 
on outflow models.\\
Finally, in Section~\ref{sec:Summary} we present our conclusions.

\section{Simulations} \label{sec:sims}
\subsection{Simulations and Sample Definition}

\begin{deluxetable}{cccccc}
\footnotesize
\tablecaption{Simulation parameters.}
\tablewidth{0pt}
\tablehead{
\colhead{$L$\tablenotemark{a}} &
\colhead{$\epsilon$\tablenotemark{b}} &
\colhead{$m_{\rm SPH}$\tablenotemark{c}} &
\colhead{$m_{\rm dark}$\tablenotemark{c}} &
\colhead{$M_{\rm *,min}$\tablenotemark{c,d}} &
\colhead{$z_{\rm end}$}
}
\startdata
$16$ & $1.25$ & $3.87$ & $25.2$ & $248$ & 2,0\tablenotemark{e} \\
$32$ & $2.5$ & $31.0$ & $201$ & $1984$ & 2,0\tablenotemark{e} \\
\enddata
\tablenotetext{a}{Box length of cubic volume, in comoving $\hmpc$.}
\tablenotetext{b}{Equivalent Plummer gravitational softening length, in comoving $\hkpc$.}
\tablenotetext{c}{All masses quoted in units of $10^6M_\odot$.}
\tablenotetext{d}{Minimum resolved galaxy stellar mass.}
\tablenotetext{e}{First number for nw, cw runs; second for vzw.}
\label{table:sims}
\end{deluxetable}

We employ the parallel cosmological galaxy formation code Gadget-2
\citep{spr02} in this study.  This code uses an entropy-conservative
formulation of smoothed particle hydrodynamics (SPH) along with a
tree-particle-mesh algorithm for handling gravity.  Heating is included
via a spatially uniform photoionizing background~\citep{haa01}.
Gas particles undergo radiative cooling under the assumption of
ionization equilibrium, where we account for metal-line cooling
using the collisional ionization equilibrium tables of \citet{sut93}.
The metal cooling function is interpolated to the gas metallicity as
tracked self-consistently by Gadget-2 \citep[for details see][]{opp06}.
Stars are formed from dense gas via a subresolution multi-phase model
that tracks condensation and evaporation in the interstellar medium
following \citet{mck77}.  The model is tuned via a single parameter, the
star formation timescale, to reproduce the~\citet{ken98} relation; see
\citet{spr03a} for details.  Star-forming gas continually self-enriches
under an instantaneous recycling approximation.  The effects of Type Ia 
supernovae are neglected; this should not affect comparisons with observed 
Oxygen abundances.  In reality Type II supernovae occur with a time delay
of 10--30 Myr, longer than our typical timestep of $\sim$a few Myr, hence 
the assumption of instantaneous feedback is inappropriate if gas accretion 
or star formation is believed to vary on shorter timescales.  However, 
our simulations (and observations;~\citealt{noe07a}) 
indicate that star formation occurs in a predominantly smooth 
fashion, hence instantaneous feedback is unlikely to introduce 
significant errors.  Delayed feedback from low-mass stars is also 
neglected.  When star particles are spawned (in two stages, each with 
half the original gas particle's mass), they inherit the metallicity of the 
parent gas particle and from then on cannot be further enriched.

We model galactic outflows using a Monte Carlo approach.  First we define
the wind model by two parameters: a wind speed $\vw$ and a mass loading
factor $\mlf$, which is the ratio of the outflow mass rate to the star
formation rate.  These parameters can be chosen to be constant or scale
with galaxy properties.  During the simulation run, for each star-forming
particle we compute a probability that it enters into an outflow based
on its star formation rate and $\mlf$, and use a random number to decide
whether that particle will enter into an outflow.  If so, we kick it
with a velocity of $\vw$ in the direction of {\bf v}$\times${\bf a},
which would be purely unipolar for a thin disk but more typically has
a large opening angle of $\sim45^\circ$.  The hydrodynamic forces are
turned off for that wind particle until it reaches a density of one-tenth
the critical density for star formation, or else it travels for a time
greater than (20~kpc/h)$/\vw$.  The outflow particle carries its own
metals out of the galaxy; it is not preferentially enriched.

In order to explore the effects of superwind feedback on the MZR, we
concentrate on three outflow schemes: A no-wind (nw) model, a ``constant
wind" (cw) model where $\vw=484\;\kms$ and $\mlf=2$~\citep[this is the
scheme used in the runs of][]{spr03b}, and a ``momentum-driven wind" (vzw) model
where $\vw\propto \sigma$ and $\eta=(300\;\kms)/\sigma$, where the
velocity dispersion $\sigma$ is estimated from the local gravitational
potential.  The exact scalings are taken from the momentum-driven wind
model of \citet{mur05}; see \citet{opp06} for details.  The vzw model
also gives a velocity boost in low-metallicity systems, based on the
arguments that more UV photons are produced per unit stellar mass at
lower metallicities \citep[specifically, we employ eqn.~1 of][]{sch03},
and that it is these UV photons that are driving the wind \citep{mur05}.
Using larger simulations evolved to low redshift, we have verified that
both models broadly reproduce the star formation history 
of the universe~\citep[][Figure 4]{opp06}; this constrains the choice 
of parameter values.

We run a total of six simulations: two different volumes, each with our 
three different superwind schemes, as detailed in Table~\ref{table:sims}.  
All runs assume a WMAP-concordant cosmology \citep{spe03} having $\Omega=0.3$,
$\Lambda=0.7$, $H_0=70\kmsmpc$, $\sigma_8=0.9$, and $\Omega_b=0.04$.
Each run has $256^3$ dark matter and $256^3$ gas particles, evolved
from well in the linear regime.  

We identify galaxies using Spline Kernel Interpolative DENMAX
and dark matter halos using the spherical overdensity algorithm
\citep[see][for full descriptions]{ker05}.  In previous papers we
have shown that imposing a mass resolution cut of 64 star particles
leads to a converged sample in terms of stellar mass and star formation
history~\citep{fin06,fin07,dav06a}.  In the present work we make an even
more conservative cut at 128 star particles in order to study the scatter
in our simulated trends.  Hence our minimum resolved galaxy stellar mass
ranges from $2.5\times 10^8\msun$ in our higher-resolution $16\hmpc$
runs to $2.0\times10^9\msun$ in our larger volume $32\hmpc$ runs.  Just as
we did in \cite{dav06a}, we will often show both runs on the same plot,
and the smoothness of trends in overlapping mass ranges is to be noted
as an indicator of numerical resolution convergence.  We have also run
higher-resolution simulations in $8\hmpc$ volumes and confirmed that the
trends that we identify continue to lower masses, though the galaxies in
these runs are unobservably small at the redshifts where we will perform
comparisons to data.

\subsection{Scales in the Wind Models} \label{ssec:scales}

Our wind models introduce several mass scales that are important in
understanding the behavior of the MZR.  The {\it reheating scale} is the
scale below which galaxies produce enough supernova energy to unbind all
of their gas.  The {\it blowout scale} is the scale below which the wind 
speed exceeds the escape velocity of the halo.  Here we compute values 
for these scales in our wind models.

We first consider the reheating scale in the cw model.  The virial energy
of the baryons in a halo scales with the halo mass as $E_{\rm vir} \propto
M^{5/3}$ whereas the feedback energy scales as $E_{\rm wind} \propto M$
(assuming that the fraction of baryons converted to stars $f_*$
varies slowly with $M$).  The ratio of these energies thus scales as
$E_{\rm wind}/E_{\rm vir} \propto M^{-2/3}$ with the implication that
the relative importance of wind heating declines as mass increases in the
cw model.  Below the ``reheating scale", a galaxy's winds produce enough
energy to expel all of the baryons from the halo; this is analogous to the
critical scale for supernova-driven mass loss proposed by~\citet{dek86}.
In the spherical collapse scenario~\citep[e.g.,][]{dek03} it can be shown
that this scale corresponds to a stellar mass of
\begin{eqnarray} \label{eqn:reheat}
\mstar & = & \frac{f_*^{5/2} \mlf^{3/2} \vw^3}{2^{3/2} G^{3/2}}
\frac{\Omega_b}{\Omega_m} \left[\frac{4\pi}{3}\Delta(z)\rho(z)\right]^{-1/2}\\
       & = & \frac{4\times10^{10}\msun}{(1+z)^{3/2}}
            \left(\frac{f_*}{0.1}\right)^{5/2} 
           \left(\frac{\mlf}{2}\right)^{3/2}
           \left(\frac{\vw}{484\; \rm{km\;s}^{-1}}\right)^3
\end{eqnarray}
or a halo circular velocity of
\begin{eqnarray} \label{eqn:blowout}
V     & = & \vw \left(\frac{f_*\mlf}{2}\right)^{1/2}\\
      & = & 150\; {\rm km\; s}^{-1} \left(\frac{\vw}{484 \rm{km\; s}^{-1}}\right)
                                 \left(\frac{f_*}{0.1}
                                 \frac{\mlf}{2}\right)^{1/2}.
\end{eqnarray}
Here, $\Delta(z)\approx200$ is the overdensity of collapsed structures, 
which varies weakly with redshift~\citep{dek03}; $\mlf$ is the mass loading 
factor of the winds; and $\vw$ is the wind speed.
Note that the circular velocity of the reheating scale in the cw model 
is similar to the critical velocity identified by~\citet{dek86}.  This 
is expected because the energy injected by winds assuming these parameters 
is comparable to the supernova feedback energy~\citep{spr03b}.  Hence a
comparison between our cw model and observations constitutes a quantitative 
test of the~\citet{dek86} scenario.

In galaxies above the blowout scale, our constant wind model outflows
should escape from the galaxies' halos provided that they couple
inefficiently with the ambient halo gas.  Under reasonable assumptions
regarding the depth of the galaxy's potential well, the ratio of the
escape velocity to the halo circular velocity lies within the range
$V_{\rm{esc}}/V \approx$ 1.5--3.5~\citep[e.g.,][]{mar99}.  In the
spherical collapse model, the blowout scale in the cw model is therefore
given by
\begin{eqnarray}
\mstar & = & \frac{f_* \vw^3 (V/V_{\rm{esc}})^3}{G^{3/2}}
	     \frac{\Omega_b}{\Omega_m}
	     \left[\frac{4\pi}{3}\Delta(z)\rho(z)\right]^{-1/2}\\
       & = & \frac{6\times10^{10}\msun}{(1+z)^{3/2}}
             \left(\frac{f_*}{0.1}\right)
           \left(\frac{\vw}{484\; \rm{km\; s^{-1}}}\frac{V/V_{\rm{esc}}}{0.4}\right)^3
\end{eqnarray}

Both the reheating scale and the blowout scale fall within the range that 
is resolved by our simulations, hence we expect to see features in the 
MZR indicating that the two effects grow significantly more effective 
below this scale and decreasingly effective above it.  In 
Section~\ref{ssec:eta} we will show that this is indeed what happens, 
although we will argue that blowout is much more important than reheating in
our models.  In Section~\ref{sec:compare_analytics} we will use simple 
physical arguments to show how the existence of a blowout scale leads to 
predictions that conflict with observations.

In the vzw model, $\mlf = \sigma_0/\sigma$ and $\vw=k\sigma$, where
$\sigma_0$ is a constant, $\sigma$ is the halo velocity dispersion,
and $k$ relates the wind velocity and the halo velocity dispersion
to the ratio of the galaxy luminosity to the galactic Eddington
luminosity~\citep{mur05}; $k=6.7$ on average.  Hence by construction,
all galaxies are above the blowout scale in this model.  With these
assumptions, the injected energy scales as $E_{\rm wind} \propto M
\mlf \sigma^2 \propto \sigma^4 \propto M^{4/3}$, and the ratio of the
heating energy to the virial energy scales as $E_{\rm wind}/E_{\rm vir}
\propto M^{-1/3}$---more shallowly than in the cw model.  In this case,
the reheating scale is given by
\begin{eqnarray}
\mstar & = & \frac{k^6\sigma_0^3 f_*^4}{8 G^{3/2}} \left(\frac{\Omega_b}{\Omega_m}\right)
             (\frac{4}{3} \pi \Delta(z) \rho(z))^{-1/2} \\
       & = & \frac{3\times10^{12}\msun}{(1+z)^{3/2}}
             \left(\frac{\sigma_0}{300 {\rm\;km\;s^{-1}}}\right)^3
             \left(\frac{f_*}{0.1}\right)^4,
\end{eqnarray}
which is so large that in practice all of our simulated galaxies would
expel all their baryons if wind energy coupled efficiently with the
remaining gas in the galaxies' halos.  In Figure~\ref{fig:eta} we will
show that this does not happen, and halos actually retain much of their
baryonic mass far below the reheating scale.  Hence in our models, the
outflow energy does not couple efficiently to the ambient ISM or halo gas,
but rather tends to blow holes and escape into the IGM.
This idea is qualitatively consistent with high-resolution individual
galaxy simulations of outflows~\citep[e.g.][]{mac99}, but inconsistent
with the assumption of efficient energy coupling with ambient gas that
is sometimes made in analytical blowout models.

\section{The $M_*-Z_g$ relation at $z=2$} \label{sec:mmr_z2}

\begin{figure}
\setlength{\epsfxsize}{0.5\textwidth}
\centerline{\epsfbox{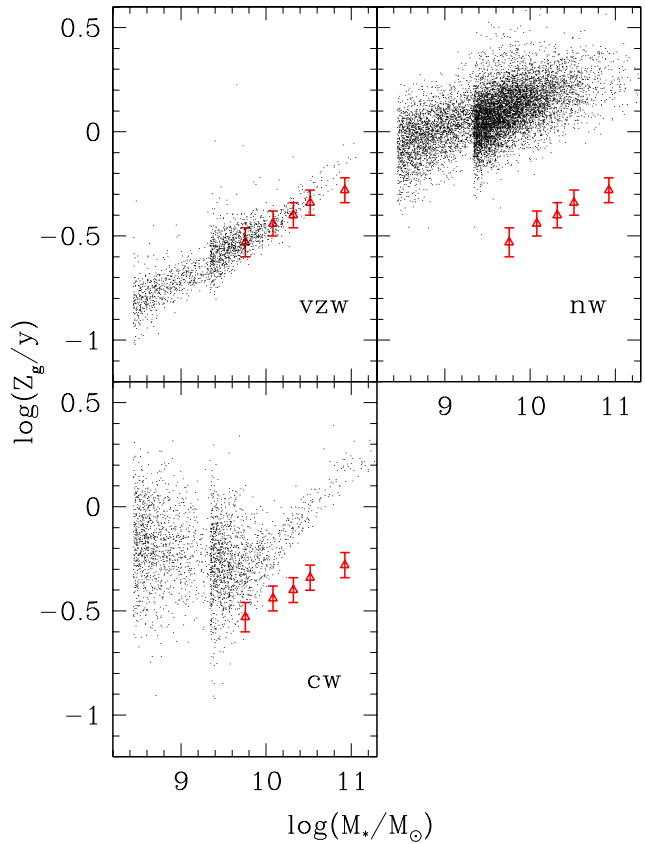}}
\vskip -0.0in
\caption{The MZRs at $z=2$ compared to
observations by~\citet{erb06} (reproduced from~\citealt{dav06b}).  
The no-wind case overproduces metals, the constant wind case shows
too steep a slope above the blowout scale and a large scatter below it,
and the momentum-driven wind scenario fits observations quite well.
The two clumps of points in each figure correspond to the $16$ and
$32\hmpc$ simulation volumes, and are bounded at the low-mass end by their 
galaxy mass resolution limits.
}
\label{fig:mmr0}
\end{figure}

\subsection{Gas-phase Metallicities}\label{mzr_g}
We begin by demonstrating that the choice of wind model heavily impacts
the simulation's predictions for the MZR.  Before doing so, several
remarks are in order regarding the simulated and observed measurements.
First, throughout this work we follow~\citet{dav06b} and define the
metallicity of each simulated galaxy as the SFR-weighted metallicity
of its gas particles; this presumably provides a fair analogue to
metallicities derived from metal emission lines as these trace the most
actively star-forming regions in each galaxy.  Second, in order to
compare our simulated metallicites with the measurements of~\citet{erb06}
in Figures~\ref{fig:mmr0} and~\ref{fig:zfg}, we normalize all metallicities 
to the net yield.  We have calculated the net Oxygen yield using 
published Type II SN yields~\citep{ww95,cl04,po98} with the~\citet{sal55} 
and~\citet{cha03} IMFs and find that it lies between 0.008 and 0.021 
depending on the choice of models, IMF, and metallicity.  Our simulations
assume a total metal yield of 0.02 with solar abundance ratios,
corresponding to a net Oxygen yield of 0.0087.  We use this value 
to normalize the~\citet{erb06} measurements for consistency with 
our simulations while noting that the choice of net yield introduces
a factor of $\sim$2 uncertainty in the relative normalizations in
Figure~\ref{fig:mmr0} that is in addition to the factor of $\sim2$
uncertainty in the observed metallicities~\citep{sha05}.  Finally, 
because the stellar masses reported in~\citet{erb06} correspond 
to the total stellar mass formed whereas our simulations report the mass 
of stars that does not immediately explode in Type II supernovae, we
multiply the~\citet{erb06} masses by the ratio of mass in long-lived
stars to the total mass formed (which is 0.802 for the~\citealt{cha03} 
IMF) while noting that the observed stellar masses are also uncertain 
at the $\sim2\times$ level owing to uncertainty in the IMF.

Figure~\ref{fig:mmr0} shows the MZR that arises in each wind model 
at $z=2$, compared with observations by~\citet{erb06}.  
The fact that outflows affect the MZR is abundantly clear
from these figures.  In the no wind case, galaxies are too enriched at
a given stellar mass~\citep[this is also seen in the wind-free models
of][]{ros06}.  This indicates that outflows are necessary to expel metals
from galaxies, adding to the growing body of evidence that outflows
have a significant and ubiquitous impact on high-redshift galaxies.
Interestingly, the slope of the MZR in the no-wind case is in fair
agreement with observations (though slightly too shallow in detail),
despite the fact that no galaxies are driving out any metals.  Metal loss
by tidal stripping is generally negligible, as we show in 
\S\ref{ssec:metret}.  Hence the slope of the MZR does {\it not} necessarily
imply that low-mass galaxies preferentially expel metals relative to
high-mass ones.  Even without outflows, the observed MZR slope (though
not its amplitude) is broadly reproduced.

The constant wind (cw) model shows some remarkable and interesting
trends.  Above the blowout scale of $10^{10}M_\odot$, the MZR slope is
quite steep, steeper than the no-wind case, and the scatter is fairly
small.  Hence relative to the no-wind case, the cw model appears to
be preferentially ejecting metals from low-mass systems, as expected.
Unfortunately, this takes what was a good agreement with the observed
slope in the nw case and produces poor agreement in the cw case.
Below the blowout scale, the scatter increases dramatically; this
behavior is seen at all redshifts down from $z\approx 6\rightarrow 0$ and
is not a consequence of numerical resolution, as evidenced by the fact
that our higher-resolution run joins smoothly onto the lower-resolution
one.  The large scatter in the cw model can be qualitatively compared
with the findings of~\citet{geh06}, who report that the observed
baryonic Tully-Fisher relation does not show the excess scatter below
the blowout scale that would be predicted in blowout scenarios.
Additionally,~\citet{lee06} have shown that the observed scatter is
$\approx0.1$ dex at all mass scales at low redshift and argued that
this is inconsistent with blowout scenarios.  We discuss the likely
source of this excess scatter in Section~\ref{ssec:compare_scatter}; 
the important point is that this scatter is not observed, hence 
models that introduce a characteristic blowout scale at observable 
masses are unlikely to be consistent with~\citet{lee06}.

The vzw model's MZR shows good agreement with observations in both slope
and amplitude.  There is a minor offset in amplitude, but there are
enough systematic uncertainties in metallicity indicators to render this
difference insignificant~\citep[e.g.][]{tre04,ell05}.  To explain this
agreement, \citet{dav06b} proposed that galaxies must lose a roughly 
fixed fraction of their metals independent of their mass, because this 
would preserve the slope seen in the nw case while lowering the 
amplitude.  Indeed, by comparing the mass of metals produced by each 
galaxy in the vzw run with the mass of metals retained in stars and gas, 
we find the fraction of metals that galaxies retain scales very slowly as 
$\mstar^{0.07}$ at $z=2$ (\S\ref{ssec:metret}).  It is intriguing
to compare this result to~\citet{geh06}, who infer that the fraction of
baryons lost cannot vary with baryonic mass based on the fact that the
slope of the observed $z\sim 0$ baryonic Tully-Fisher relation lies very
close to the value that is expected from cosmological simulations that
do not treat baryons.  In the momentum-driven wind scenario this 
arises naturally because outflow speeds scale with galaxy escape 
velocities.

\subsection{Stellar Metallicities}\label{mzr_s}
\begin{figure}
\setlength{\epsfxsize}{0.5\textwidth}
\centerline{\epsfbox{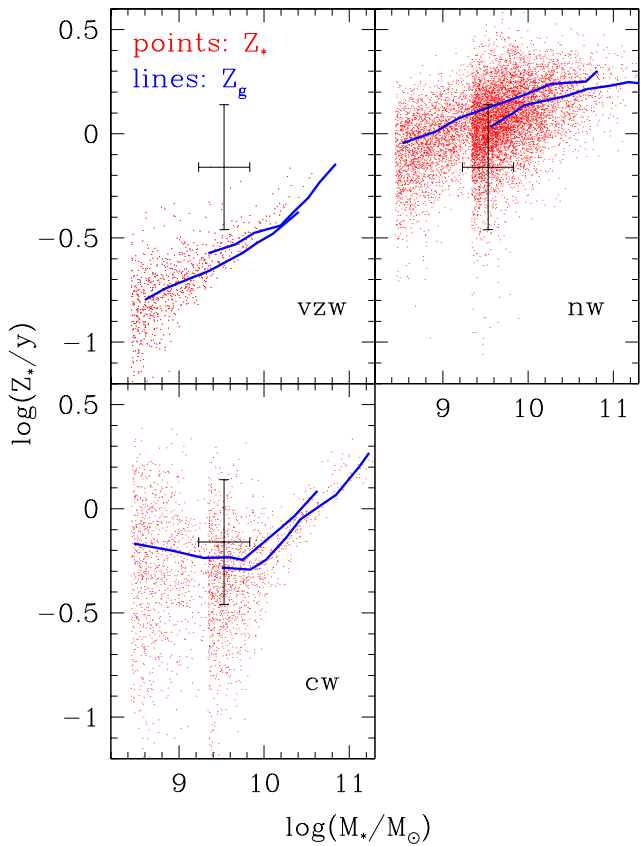}}
\vskip -0.0in
\caption{$Z_*$ versus $M_*$ at $z=2$.  Red dots denote galaxies in our
simulations while blue curves denote the mean trend from the gas-phase MZR
(Figure~\ref{fig:mmr0}).  The data point with error bars denotes the 
inferred stellar mass and metallicity of MS 1512-cB58.  Within each
wind model, the UV-weighted stellar metallicities closely track the 
metallicity of the star-forming gas, as expected.
}
\label{fig:mzrs}
\end{figure}
Although the goal of this paper is to understand the relationship 
between stellar mass and gas-phase metallicity, recent observational
efforts to constrain the evolution of the stellar mass-stellar metallicity
relation (hereafter, the ``stellar MZR") motivate a comparison between 
our simulated gas-phase and stellar metallicities.  Before we proceed, 
however, a few comments on systematic effects are in order.  The first 
observational constraints on the high-redshift stellar MZR will focus on 
rest-frame UV absorption features~\citep[e.g., the 1978 index of][]{rix04} 
because at $z\geq2$ these features are redshifted into conveniently-accessible 
optical wavelengths.  Unfortunately, they are also dominated by stars with 
zero-age main-sequence masses greater than $5 M_\odot$~\citep{rix04},
hence they are only sensitive to a galaxies' youngest stars.  In order 
to estimate the extent to which the observed stellar MZR depends on the 
choice of metallicity indicator, we have measured the stellar MZR for 
each wind model twice: once using mass-weighted stellar metallicities 
averaged over all stars ($\zsall$), and once using only stars that are
younger than 100 Myr ($\zsuv$).  The latter figure approximates a 
UV-luminosity-weighted stellar metallicity.  We find that, for all 
three wind models, $\zsuv$ lies systematically 40--60\% higher 
than $\zsall$, indicating that UV indices systematically 
overestimate stellar populations' mean metallicities even in the 
absence of any systematics in the indices themselves.  We also 
find that, in all three wind models, the scatter in $\zsuv$ at 
given $M_*$ is 0.1--0.2 dex, whereas the scatter in $\zsall$ 
is generally $\approx0.05$ dex, tighter than the scatter in the 
gas-phase MZR.  The increased scatter in $\zsuv$ likely owes 
partially to our simulations' limited mass resolution.  On the other 
hand, it is also likely that the short time baseline sampled by UV 
indices renders them more sensitive to the distribution of 
metallicities among individual HII regions within individual galaxies,
hence we do expect the scatter in the measured MZR to be larger 
for UV indices than for optical indices.  In order to facilitate 
comparison with upcoming measurements at high redshift, we only
consider $\zsuv$ throughout the rest of this work.

The red points in Figure~\ref{fig:mzrs} compare the predicted stellar MZRs
at $z=2$ for our three wind models (red points) to the mean gas-phase MZRs
(blue curves) as well as to current constraints on the $z=2.7276$ lensed 
galaxy MS 1512-cB58 (hereafter, ``cB58";\citealt{yee96}).  For cB58, we 
use the stellar metallicity of~\citet{rix04} normalized to solar yield and 
the stellar mass of~\citet{bak04}.  We assume a factor of 2 uncertainty 
for each estimate.  Comparing the stellar and gas-phase MZRs in 
Figure~\ref{fig:mzrs} indicates that they are predicted to be quite similar, 
regardless of the outflow scenario.  This is expected since young stars' 
metallicities should track the metallicities of their parent gas clouds.  
On the other hand, it is not consistent with observations of cB58, for 
which the most likely stellar metallicity $Z_*$=0.7$Z_\odot$ significantly
exceeds the inferred gas-phase metallicity 
$Z_g=0.4\pm0.1Z_\odot$~\citep{rix04}.  However, considering that the 
metallicity offset in cB58 is well within the range of systematic 
uncertainties and that its ISM metallicity shows excellent agreement with 
the observed mean gas-phase metallicity for its stellar mass at 
$z\sim2$~\citep{erb06}, we believe that the offset likely results from 
systematic offsets in the observational abundance indicators.  It will be 
interesting to see whether larger samples of galaxies at high redshift show 
a similar offset.

We have additionally compared our stellar MZRs to the stellar MZR predicted 
at $z=2$ by the hypernova feedback model of~\citet[][Figure 20]{kob07}.  In 
the range $10^{9-10.5}\msun$, the~\citet{kob07} model predicts V-band 
luminosity-weighted stellar metallicities that are $\geq 0.5$ dex below our 
nw model at all scales, 0.1-0.2 dex higher than our vzw model at all scales, 
and 0.2--0.5 dex below our cw model at all masses except the blowout scale.  
The best agreement is hence with our favored vzw model, indicating that, 
broadly, the effects of the~\citet{kob07} hypernova feedback model are 
similar to our momentum-driven outflows although the hypernova winds may 
be somewhat weaker.

\subsection{The IGM Metallicity}\label{mzr_igm}
A fully self-consistent model for galaxy evolution must account for the 
distribution of metals in the IGM as well as in galaxies, hence we have 
also tested our outflow models by comparing the predicted and observed 
IGM metallicities.  Because the impact of our outflow prescriptions on 
the distribution of metals in different gas phases has been discussed 
previously~\citep{opp06,dav07}, we will only mention the results here.  
The mean metallicity of the IGM at $z=2$ in our favored vzw model is 
roughly [Z/H]=-1.7~\citep[][Figure 1]{dav07}.  This is roughly twice 
as high as predicted by the hypernova feedback model 
of~\citet[][Figure 18]{kob07}, implying once again that the outflows in
the hypernova feedback model are somewhat weaker.  However, by analyzing 
simulated quasar absorption spectra generated along sightlines through the 
simulation volume,~\citet{opp06} have shown that the predicted abundance 
of CIV in the vzw model is in excellent agreement with observations from
$z=6\rightarrow2$.  Hence our favored vzw model is broadly consistent with
observations of the distribution of metals both within and 
outside of galaxies at $z=2$.

\section{Analytical Model for the MZR} \label{sec:analytics}

Figure~\ref{fig:mmr0} strongly suggests that the observed MZR can
be used to constrain superwind models, but it gives little physical
insight into how superwinds impact the trends and evolutionary behavior
of the MZR.  In order to probe this question more deeply, we construct
a simple one-zone chemical enrichment model similar to many others in
the literature~\citep[e.g.,][]{tin80}.  The novel aspect is that we
will use our simulations to calibrate the model inputs, so essentially by
construction our analytical model will broadly reproduce the simulation
results, as we show in \S\ref{ssec:compare_evol}.  Owing to its
simplicity, it provides an instructive tool to examine the relative
importance of various physical effects in driving the MZR, which we will
do in \S\ref{sec:compare_analytics}.

\subsection{Equations of Evolution} \label{ssec:eq_of_ev}
At each timestep, the mass of metals in the ISM $M_{Z}$ increases owing 
to inflows and star formation and decreases owing to outflows and metals 
being locked up in long-lived stars:
\begin{eqnarray}
\dot{M}_{Z} & = & Z_{\rm{IGM}}\acc + y\sfr - 
Z_g\sfr - Z_g\dot{M}_{\rm{wind}} \nonumber \\
 & = & \alpha_ZZ_g\acc + y\sfr - 
Z_g\sfr(1+\mlf) \label{eqn:evol_Mz}
\end{eqnarray}
Here, $Z_{\rm{IGM}}$ and $Z_g$ denote the metallicities of the IGM and the
ISM, respectively; $\acc$ and $\sfr$ denote the gas accretion and star
formation rates, respectively; $y$ denotes the net Oxygen yield~\citep{tin80},
$\dot{M}_{\rm{wind}}$ denotes the rate at which gas enters the wind; and
$\mlf$ denotes the ratio $\dot{M}_{\rm{wind}}/\sfr$.  
Equation~\ref{eqn:evol_Mz} makes the following assumptions:
\begin{itemize}
\item The rate at which new metals are injected into the ISM is given by 
$y\sfr$, where the yield $y$ is a constant;
\item The outflow rate is proportional to $\sfr$;
\item The metallicity of the outflowing gas is equal to the mean metallicity 
of the galaxy's ISM;
\item We assume instantaneous feedback~\citep{tin80} so that the effects of 
Type Ia supernovae and delayed mass loss are neglected;
\item The mean metallicity of inflowing gas is some fraction $\alpha_Z\geq0$ 
of the metallicity in the galaxy's ISM; 
\item We neglect the difference between the rate of formation of all stars
and the rate of formation of low-mass stars, i.e. we assume that the bulk of
stellar mass is in long-lived stars; and
\item All metals resulting from star formation and gas accretion are assumed
to be well-mixed within the galaxy's ISM.
\end{itemize}
The first four assumptions mimic those made in our simulations whereas the 
latter three are made for convenience.  Also, note that any effects 
resulting from the enhanced cooling rates of metal-enriched gas will be 
taken into account implicitly when we tune our star formation efficiencies 
to match the simulations in Section~\ref{ssec:etastar}.

To obtain the metallicity evolution in this model we need the evolution
of metal mass as a function of gas mass.  This requires us to know
all the parameters in equation~\ref{eqn:evol_Mz} except the net yield
$y$ (we normalize our metallicities by $y$ in all of our Figures).
We will assume $\alpha_Z=0$ for the nw and cw models and 50\% for the
vzw model (we will justify this assumption in \S~\ref{ssec:ns_wind}).
For $\mlf$ in the vzw model, we infer $\sigma$ from the time-dependent 
relation between baryonic mass and velocity dispersion that we measure 
directly from the vzw outputs and then plug this into 
$\mlf = 300\kms/\sigma$.  For the cw model, we set $\mlf=(2,0)$ for all 
galaxies with masses (below, above) the blowout scale.

Finally, we need $\acc$ and $\sfr$ as a function of redshift and galaxy
mass.  In the following sections we describe how we calibrate relations
for these quantities from our simulations.

\subsection{Gas Accretion History} \label{ssec:acchist}

\begin{figure}
\setlength{\epsfxsize}{0.5\textwidth}
\centerline{\epsfbox{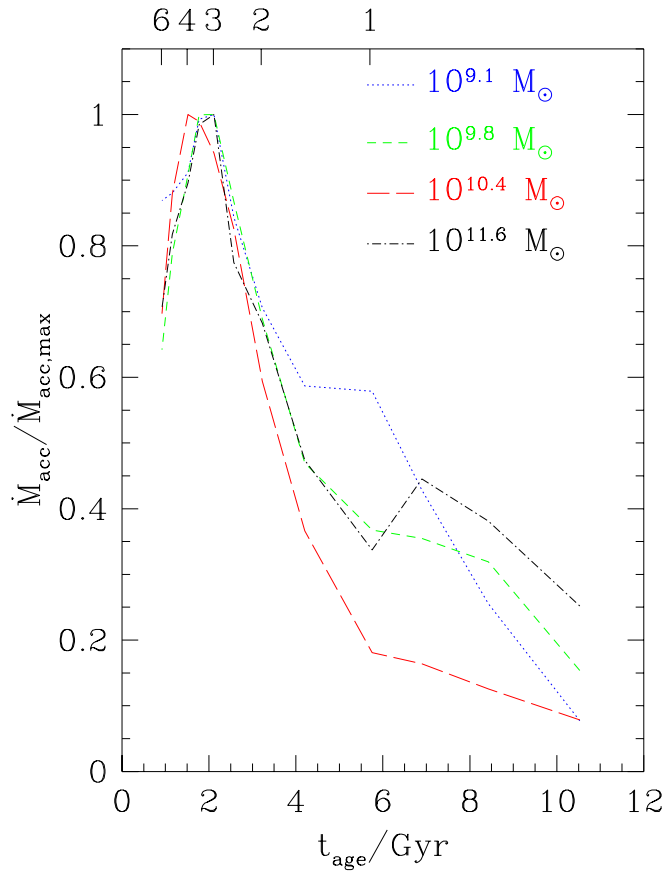}}
\vskip -0.0in
\caption{Mean baryonic accretion rates of the simulated galaxies in 
the $32\hmpc$ vzw simulation as a function of time (bottom) and redshift 
(top).  Four different mass bins are shown, with the labels referring to 
the stellar mass at $z=0$; each history is normalized to its maximum 
accretion rate.  Galaxies of all masses experience remarkably
similar baryonic accretion histories.
}
\label{fig:acchist}
\end{figure}

We infer the time-dependent rate at which galaxies accrete fresh gas from
their environments directly from the vzw simulations.  In detail, we trace
each galaxy's baryonic growth rate backwards in time by searching for its
most massive progenitor in each simulation snapshot.  The total baryonic 
accretion rate is then simply the baryonic growth rate plus losses due to 
outflows:
\begin{equation}
\acc = \dot{M}_{\rm{bar}} + \mlf \sfr.
\end{equation}
In practice this technique yields the accretion rate of gas and stars
rather than just gas as it is difficult to determine from the outputs
of our simulations whether new stellar mass results from star formation
or mergers.  However, baryonic mass growth is known to be dominated by
gas accretion at high redshifts~\citep{ker05,gw07}, and the bottom panel
of Figure~\ref{fig:Zg_eq} suggests that this is approximately true in our
models as well.

To estimate outflow losses for the vzw model, we need the average
$\mlf$ for each galaxy as a function of time.  We can obtain a good
estimate by combining the known baryonic mass with the time-dependent
relation between galaxy velocity dispersion and baryonic mass,
measured directly from our simulations, and substituting this back into
the relation assumed by our vzw model $\mlf=(300\kms/\sigma)$.
Figure~\ref{fig:acchist} plots the resulting accretion histories in 4
bins of stellar mass, where we have normalized each one to its maximum
accretion rate; the mean stellar masses in each bin at $z=2$ are
indicated.

Galaxies of all masses experience remarkably similar accretion histories
in our simulations, with accretion rates rising steadily at early times
until a peak around $z\sim$4--3 and falling afterwards.  The existence
of this generic accretion history has two interesting implications.
First, under the assumption that $\sfr$ tracks $\acc$ (which we justify
below), the rising accretion rates immediately explain why galaxies
generically exhibit rising star formation histories at high redshifts
in our simulations~\citep{fin07}.  Second, the existence of a generic
accretion history implies that a correlation between $\sfr$ and halo
mass---and, under reasonable assumptions, stellar mass---is expected,
as found in other simulations by~\citet{fin06} and recently observed
at $z\sim 0.5-2$ (\citealt{noe07a}; E.\ Daddi, private communication).

In detail, there are slight differences after $z=2$ with a suggestion 
that the most massive galaxies continue to accrete too much gas at late 
times.  Future work incorporating a treatment for AGN feedback is expected
to alleviate this problem.  Additionally, note that the use of a completely 
generic gas accretion history along with the assumption that the gas 
processing rate exactly tracks the gas accretion rate does not reproduce 
the so-called ``downsizing" of galaxy evolution, where the latter is 
defined as the tendency for more massive galaxies to exhibit lower specific 
star formation rates~\citep{zhe07,igl07}, unless the ratio of past-averaged
to present wind suppression factors $(1+\emlf)/(1+\mlf)$ scales strongly
with mass.  It is more likely star formation is delayed in low-mass 
galaxies by an effect that our model does not account for~\citep{noe07b}.

We averaged over the normalized accretion histories in
Figure~\ref{fig:acchist} and found that a reasonable approximation to
the resulting mean accretion history is given by the fitting formula
from~\citet{spr03b}:
\begin{equation}
\acc(z)  \propto \frac{b\exp[a(z-z_m)]}{b-a+a\exp[b(z-z_m)]}.
\label{eqn:acchist}
\end{equation}
Using a least-squares algorithm we determine the best-fit parameters 
for the vzw model as $a=1.06$, $b=1.32$, and $z_m=3.5$.  Although 
this formula was originally designed to describe the global star formation 
history, as \citet{ker05} has pointed out, this closely tracks the gas 
accretion history, so it is not surprising that a good fit is obtained for 
$\acc$ using this form.  

Similar trends are seen in the cw and nw models at $z\geq2$ 
(unfortunately only our vzw simulations were evolved to $z=0$).  In 
detail, there are slight differences between the gas accretion histories 
in the three wind models.  The most prominent of these is that gas 
accretion rates peak at an earlier redshift in the cw model than in 
the vzw or nw models because the cw's energetic winds heat the IGM too 
efficiently~\citep{opp06}.  As a result, the gas cooling time at moderate 
overdensities ($\sim10$) surpasses a hubble time and the star formation 
rate density begins to decline at an earlier redshift 
($z\sim5$;~\citealt{spr03b}) than in the vzw or nw models 
($z\sim3$;~\citealt{opp06}).  We account for this early peak by setting 
$z_m=5$ when modeling the impact of cw outflows in our analytical model.  
We will show in Figure~\ref{fig:hydro_analytic} that using this parameterized 
fit to the vzw model's gas accretion history allows us to approximate the
chemical evolution of our simulated galaxies in all three wind models, 
hence a more detailed discussion of the impact of outflows on gas accretion 
histories is beyond the scope of the present work.  The amplitude of a 
galaxy's gas accretion history is determined by a constant multiplicative 
factor.

\subsection{Star Formation Rates} \label{ssec:sfr}

\begin{figure}
\setlength{\epsfxsize}{0.5\textwidth}
\centerline{\epsfbox{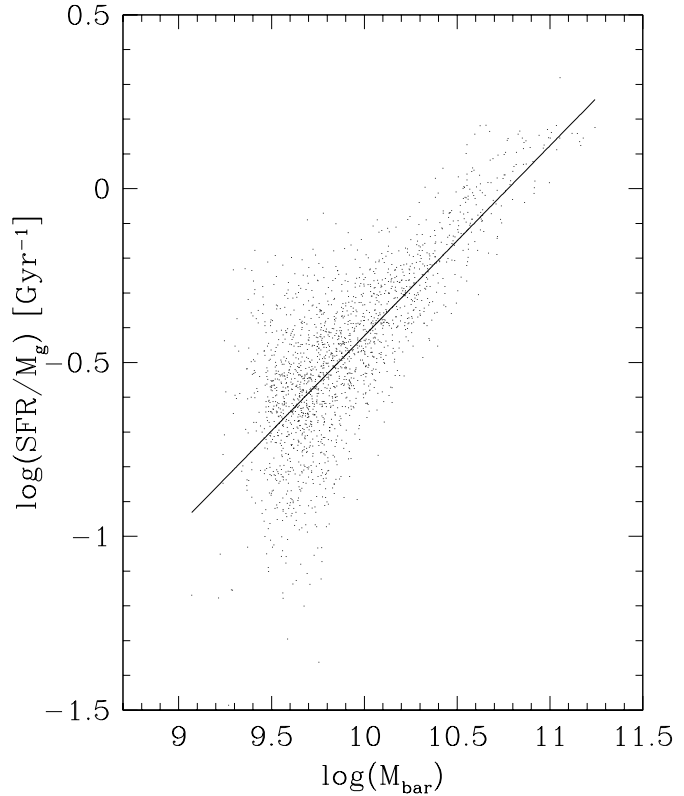}}
\vskip -0.0in
\caption{Least-squares fit to the vzw star formation efficiency at
$z=2$.  Because the slope of the trend does not vary with scale, we can 
readily use it to tune our analytical model.
}
\label{fig:sample_sfeff}
\end{figure}

\begin{figure}
\setlength{\epsfxsize}{0.5\textwidth}
\centerline{\epsfbox{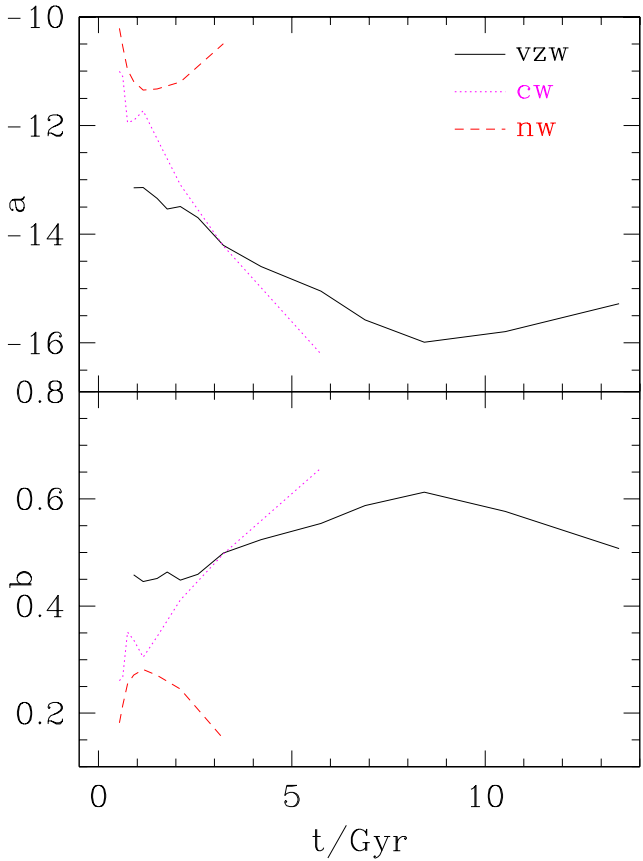}}
\vskip -0.0in
\caption{Least-squares fitting parameters to the star formation efficiency
as a function of baryonic mass in our three wind models.  For each model,
the line indicates how fits to the relation
$\log(\sfr/M_g) = a + b\log(M_{\rm{bar}})$ evolve over cosmic time.
}
\label{fig:simfits}
\end{figure}

To obtain $\sfr$, we measure $\sfr/M_g$ in the simulations as a function of
baryonic mass and redshift, and then use
\begin{eqnarray}
\sfr & = & \frac{\sfr}{M_g}(z,M_{\rm{bar}})\;M_g,
\end{eqnarray}
where the gas mass $M_g$ increases owing to inflows and decreases owing to
outflows and star formation according to
\begin{eqnarray}
\dot{M}_g & = & \acc - \sfr(1+\mlf). 
\label{eqn:evol_Mg}
\end{eqnarray}
Through trial and error we determined that it is not possible to reproduce
the detailed mass-metallicity evolution of our simulated galaxies without
such a calibration for each wind model; indeed, this is a hint as to
what governs the MZR.

Figure~\ref{fig:sample_sfeff} shows that the star formation efficiency
varies with scale in a simple way in the vzw model at $z=2$; similar
trends hold for all epochs and wind models.  In order to tune our
analytical model, we fit regression lines of the form $\log(\sfr/M_g)
= a + b\log(M_{\rm{bar}})$ to each simulation at a range of epochs.
Figure~\ref{fig:simfits} gives the resulting fits.  In all three wind
models~$b>0$, indicating that more massive galaxies are generically more
efficient at converting their gas into stars owing to their higher gas
densities; this behavior is qualitatively similar to the idea that star
formation timescales are shorter in more massive galaxies~\citep{noe07b}.
The offsets $a$ in the wind models are everywhere lower than in the nw
model because superwinds decrease galaxies' gas densities.  The slopes
$b$ are everywhere higher in the wind models because wind effectiveness
generically scales with mass.  However, the fact that the cw efficiencies
lie very close to the nw efficiencies at high redshift and then diverge
from them suggests that cw winds are relatively ineffective in the
low-mass galaxies that dominate at early times.  The fact that the
slope and offset of the trend evolve in opposite directions regardless
of wind model indicates that gas densities in low-mass galaxies are
more sensitive to changes in environment than in more massive galaxies.
It is intriguing that the slope of the evolutionary trend changes sign
near the point $z=0.5$, likely a consequence of the universe becoming
$\Lambda$-dominated.

Does self-enrichment impact our simulated star formation efficiencies?
If gas in more massive halos cools more efficiently owing to enhanced
metallicities then a scale-dependent ``positive feedback" could obtain 
between the gas cooling and self-enrichment rates; in this case the 
form of the MZR could be influenced by the relative effectiveness of 
this positive feedback cycle as a function of mass.  Comparing 
Figures~\ref{fig:mmr0} and~\ref{fig:simfits} shows that, indeed, the
nw model exhibits both the highest star formation efficiencies and
the highest metallicities, in qualitative agreement with this picture.

However, the low star formation efficiencies in our wind models can 
also be explained by the tendency for gas particles to be ejected by 
winds before their densities (and hence star formation rates) grow 
comparable to the typical densities in the nw model.  Moreover, while 
the metallicities in the nw model scale more steeply than the vzw 
model, the star formation efficiencies scale more steeply in the vzw 
model, in qualitative disagreement with the self-enrichment picture.
Additionally, we note that galaxy growth at these redshift and mass 
scales is dominated by cold-mode gas accretion~\citep{ker05}, with the
result that the gas cooling timescale is much shorter than the
dynamical timescale irrespective of its metallicity.  In 
Section~\ref{sec:compare_analytics} we will show that, in the outflow 
model that reproduces observations, the MZR can be understood entirely 
in terms of the effects of outflows.  Hence while a self-enrichment 
feedback cycle must be operating on some level, its effects on the 
observable MZR are likely to be weak compared to the effects of outflows.

\section{The Effects of Winds} \label{sec:sfefficiency}

In this section we begin our exploration of the impact of outflows on
the MZR by evaluating how winds affect our simulated galaxies; in 
short, we wish to determine what outflows {\it do} to galaxies.
In our analytical model we assume that the primary parameters through
which winds modulate the observable MZR are the star formation
efficiency, $\mlf$, and the gas accretion history.  We would like to
compare how each of these parameters scales with mass in our three
wind models.  The gas accretion history has been explored elsewhere
\citep[e.g.][]{opp06,dav06a}, and is reasonably well described by
equation~\ref{eqn:acchist}.  Hence we first discuss the impact of 
outflows on $\mlf$ and star formation efficiency.  Next, we show 
that the fraction of metals retained by galaxies does not drive 
the MZR even though it is affected by outflows.  Afterwards
we compare how outflows suppress stellar mass and gas-phase metallicity
on a galaxy-by-galaxy basis.  Finally, we discuss how winds impact the
trajectories that galaxies follow through the MZR.

\begin{figure}
\setlength{\epsfxsize}{0.5\textwidth}
\centerline{\epsfbox{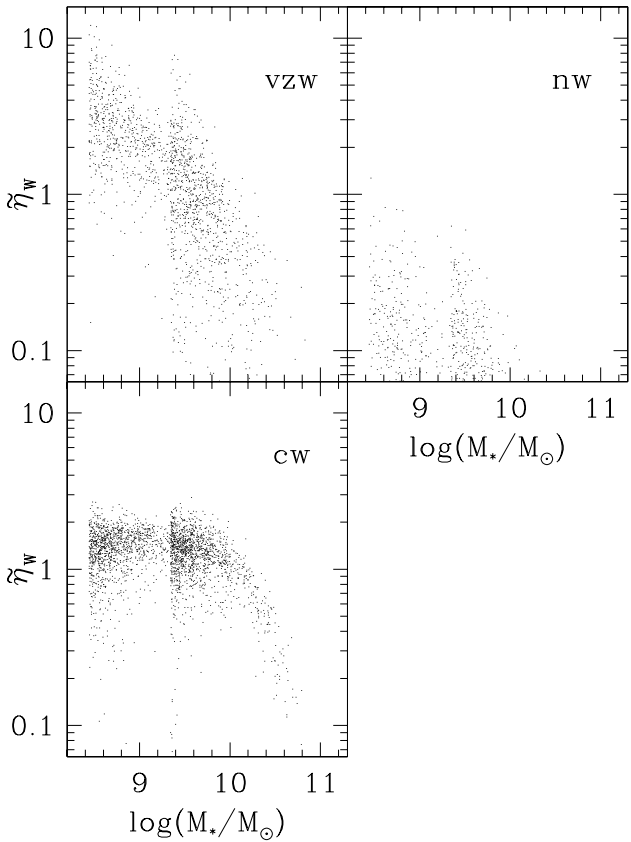}}
\vskip -0.0in
\caption{Mean mass loading factor experienced by galaxies as a function of
stellar mass.  In the nw model only low-mass galaxies lose any baryons, while
in the cw and vzw models galaxies the mass of baryons lost scale as expected.
Note how rapidly winds become ineffective above the blowout scale in the cw
model.
}
\label{fig:eta_eff}
\end{figure}

\subsection{Mass Loading} \label{ssec:eta}

In our simulations, the rate at which material enters the wind is given 
by $\sfr\mlf$.  In the cw model, $\mlf = 2$ and $\vw = 484\kms$ while in 
the vzw model $\mlf = 300\kms/\sigma$ and $\vw \approx 6.7\sigma$.
After a wind particle leaves the star forming region it interacts with
the galaxy's halo hydrodynamically.  Some gas particles may escape into
the IGM while others may radiate away their kinetic energy and rain back
down as ``galactic fountains".  Hence it is not obvious what fraction
of the gas that enters a galaxy's wind will actually escape from the
galaxy permanently, or whether this fraction will preserve the input
$\mlf$ scaling.  Fortunately, the connection between star formation,
metal enrichment, and winds allows us to constrain this quantity.

If an isolated star-forming galaxy generates a wind with a constant 
mass loading factor $\mlf$ then the fraction of metals that it 
retains is given by
\begin{equation} \label{eqn:met_retained}
\frac{M_{Z,\rm{retained}}}{M_{Z,\rm{formed}}} = 
1 - \frac{\mlf}{1+\mlf}\left(1 - \frac{Z_g M_g}{y M_*}\right).
\end{equation}
If the masses and metallicities of a galaxy's gas and stellar phases
are known then Equation~\ref{eqn:met_retained} can be solved for the
instantaneous mass-loading factor $\mlf$.  In general, $\mlf$ varies
as the galaxy grows so that it is not possible to recover $\mlf$ from
our simulation outputs.  However, in this case it is still possible
to obtain an ``effective mass loading factor" $\emlf$ as a function of
stellar mass in our simulations from Equation~\ref{eqn:met_retained}
using the known masses and metallicities of the simulated galaxies;
in this case $\emlf$ measures the mass loss rate averaged over the 
galaxy's star formation history.

As previously noted by \citet{dal06}, the fact that
equation~\ref{eqn:met_retained} is not in general proportional to the
effective yield indicates that the relative metal contribution to the
IGM from different galaxies cannot straightforwardly be inferred from
their effective yields~\citep[as attempted by, e.g.][]{bou07}.

Figure~\ref{fig:eta_eff} shows $\emlf$ vs. $M_*$ in our various wind
models at $z=2$.  This plot is one of the most important ones in this
paper for understanding the MZR.  Looking at the nw model first, we 
find that $\approx10\%$ of the nw galaxies show evidence of having lost 
some of their baryons ($\emlf > 0$).  These losses owe primarily to tidal 
stripping, although additional scatter is introduced by discreteness 
effects and uncertainties in the identification of low-mass galaxies within
the simulation outputs.  The result is a slight ``flaring" of the trends
toward low masses that is visible in all three of our models.  Because 
these effects are small compared to the overall trends that relate to
the MZR (for example, compare the size of the $\emlf$ in the nw model 
to the typical $\emlf$ in the vzw model), they do not affect any of 
our results.

In the cw model, below the blowout scale 
$\emlf\approx 1.5 \approx \frac{3}{4}\mlf$, indicating 
that on average $\frac{3}{4}$ of wind particles from these 
galaxies leave permanently.  Above the blowout scale 
($\approx 10^{10}\msun$ at $z\sim 2-3$) $\emlf$ declines rapidly because 
these winds thermalize their kinetic energy efficiently and return to the
source galaxy as galactic fountains.  Such a phenomenon, if real, would
leave clear signatures in the observed baryonic Tully-Fisher relation
or MZR at this scale.  As we discuss in \S\ref{sec:compare_analytics},
the absence of such features in observations argues against the cw model.

In the vzw model, $\emlf\propto \mstar^{-0.25}$, shallower than the
predicted slope of $-1/3$ if the stellar mass is a fixed fraction of
the halo mass.  To understand this, in Figure~\ref{fig:eta} we show
the stellar fraction $f_*$ as a function of $M_{\rm{halo}}$, and
indeed we see that for the vzw case $f_*\propto M_{\rm{halo}}^{1/3}$.
Taking this into account, we find that $\emlf\propto M_{\rm{halo}}^{-1/3}
\propto 1/\sigma$ as expected.  This indicates that outflow processes
preserve the assumed scalings once the scaling of $f_*$ is accounted
for as long as the galaxy is above the blowout scale.  Furthermore, it
indicates that in the vzw model the fraction of wind particles
that escape the galaxy is roughly constant for all galaxy masses, as
required by observations of the low-redshift baryonic Tully-Fisher 
Relation~\citep{geh06}.

It is interesting to note that the $\emlf$ trend in our vzw scenario
is qualitatively similar to the trend of wind strength versus
mass in Figure 16 of~\citet{kob07}.  In our work, this trend results
directly from the assumed scaling of the instantaneous $\mlf$ while
in~\citet{kob07} it results from their treatment for pressure-driven
outflows from supernova and hypernova feedback.

\subsection{Star Formation Efficiency} \label{ssec:etastar}
\begin{figure}
\setlength{\epsfxsize}{0.5\textwidth}
\centerline{\epsfbox{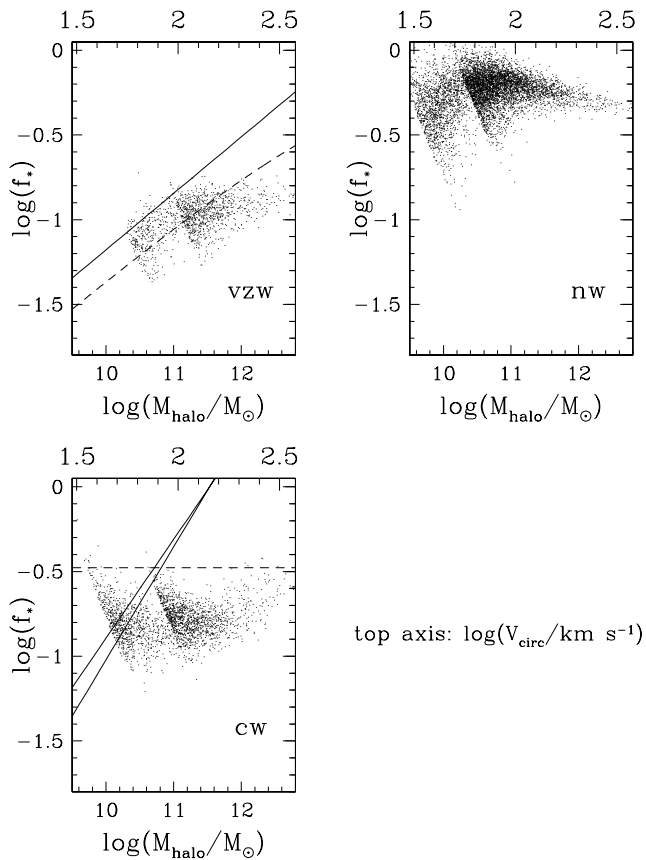}}
\vskip -0.0in
\caption{Fraction of baryons converted to stars as a function of halo mass
(bottom axis) and circular velocity (top axis) at $z=2$.  The two loci 
correspond to halos from the 16 and $32\hmpc$ volumes that contain more 
than 128 star particles.  The dashed and solid lines in the wind models 
show the range of scalings expected from 
considerations of mass-loading and energy balance, 
respectively~\citep[][see text]{dek03}.  Below the minimum mass scale
for hot-mode gas accretion $f_*$ scales with mass even without outflows.
In the presence of outflows, $f_*$ is determined by the combined effects
of mass-loading and~\citet{sca01} suppression.
}
\label{fig:eta}
\end{figure}

We now discuss the impact of outflows on the star formation efficiency.
We seek to answer two questions: (1) What impact do outflows have on
the integrated star formation efficiency $f_*$ (defined as the fraction
of baryons in a halo converted to stars as a function of halo mass) at
$z=2$?; and (2) How do outflows suppress $f_*$? The latter question is of
interest because the expected scalings depend on whether outflows couple
efficiently with the remaining gas in the galaxy's ISM as well as its
ambient IGM.  If coupling is poor then, for galaxies above the blowout
scale, the $f_*$ scaling should reflect the intrinsic $\mlf$ scaling.
Alternatively, if coupling is efficient then the $f_*$ scaling should
reflect the condition that the feedback energy is comparable to the
binding energy of the baryons in the halo~\citep[e.g.,][]{dek86,dek03}.

Figure~\ref{fig:eta} (similar to Figure 5 of~\citealt{dav06a}) shows
how $f_*$ varies with stellar mass in the different wind models at $z=2$.
Examining the nw model first, we see that in the absence of winds
$f_*$ climbs steadily with mass below $M_{\rm{halo}}=10^{11} \msun$ and
then decreases slowly with increasing mass above the characteristic minimum
mass scale for halos to be dominated by hot gas rather than cold gas 
\citep[e.g.,][]{bir07}.  This shape qualitatively mimics the behavior 
predicted by~\citet{dek03}.  However, since there are no winds to couple 
the feedback energy to the halo gas, the qualitative agreement is merely a
coincidence.  The fact that $f_*$ is not constant with stellar mass in 
the nw model has the important implication that $f_*$ (and hence the MZR)
is not governed solely by outflows~\citep{dal06}.  We note 
that~\citet[][Figure 2]{tas06} and~\citet[][Figure 2]{ros06} have observed
qualitatively similar behavior in the absence of strong outflows.

What determines the $f_*$ scaling in the absence of outflows?  If all
halos converted their gas into stars with the same instantaneous star
formation efficiency $\sfr/M_g$ then neither $Z_g$ nor $f_*$ would scale
with mass.  Evidently $\sfr/M_g$ must scale with mass at some point
prior to $z=2$.  Indeed, we find that, without outflows, star formation 
efficiency and gas density increase strongly with increasing mass before 
the reionization epoch $z\geq6$ although both trends weaken significantly 
by $z=2$.  The scaling in star formation efficiency at high redshift can 
be understood intuitively as a consequence of the fact that more 
massive halos begin collapsing earlier than less massive halos, 
giving them a ``head-start" in condensing their gas reservoirs.
The same effect also manifests itself in a trend for more massive galaxies
to exhibit older mean stellar ages than less massive galaxies in this
model, dubbed ``natural downsizing" by~\citet{nei06}.  As a result,
low-mass galaxies possess lower gas densities, enhanced gas fractions
and suppressed gas-phase metallicities with respect to massive galaxies.

The vzw model qualitatively resembles the nw model in $f_*$
vs. $M_{\rm{halo}}$ except that it is shifted down by a factor of
5--12 with an additional dependence on mass (as shown more clearly in
Figure~\ref{fig:matchGals}).  The normalization is lower because the
vzw model significantly delays star formation (and suppresses gas-phase
metallicities) at all scales.  The flattening behavior in halos above
$10^{11.5}\msun$ obtains because more massive halos are generally hot
mode-dominated, just as in the nw model.  

What determines the slope at masses below this scale? If we assume that,
as the galaxy grows, the fraction of baryons that forms stars at each
mass scale is $1/(1+\mlf)$, then $f_*$ is simply given by
\begin{eqnarray}
f_* & = & \frac{\int \rm{d}M/(1+\mlf)}{\int \rm{d}M}.
\end{eqnarray}
This scaling is denoted by the dashed line in the figure, where we have 
made the approximation that the halo velocity dispersion is given by 
$\sigma=0.0083 (M/\msun)^{1/3} \kms$.  Alternatively, if we
follow~\citet{dek03} and assume that stars continue to form until the 
total energy in outflows equals the virial energy of the halo's baryons 
then we find that $f_* \propto M^{1/3}(1+z)^{1/2}$ in the vzw model; 
this scaling is denoted by the solid line in the figure and has been 
normalized to 1 at the reheating scale at this redshift 
$M=5.6\times10^{13}\msun$.  At a glance the $\mlf$-driven explanation
seems more accurate although both theories produce the correct scaling.
However, noting that the normalizations of these two scaling relations 
are somewhat uncertain owing to the assumptions in our spherical 
collapse estimates, in practice either explanation 
could be valid.  Below we show that 
the $f_*$ scaling that is expected from energy considerations is 
not obeyed in the cw model, implying that energy in outflows does not
couple efficiently with inflowing material.  For this reason we conclude 
that the accuracy of the $f_*$ scaling that is predicted from energy 
considerations is purely a coincidence and that $f_*$ is dominated by 
the scaling of the mass loading factor $\mlf$.

In the cw model $f_*$ is relatively flat between the smallest halos
containing $\geq 128$ star particles and the blowout scale at a halo
mass of $10^{11.5} \msun$.  Above the blowout scale $f_*$ climbs slowly.
The way that $f_*$ varies only slowly with mass below the blowout scale
and climbs above it echoes the cw model's MZR (Figure~\ref{fig:mmr0}),
indicating a connection between the suppressed star formation
efficiencies and the suppressed metallicities.  However, the fact
that Figures~\ref{fig:mmr0} and~\ref{fig:eta}~look rather different in
the cw model (especially above the blowout scale) indicates that the
suppressed $f_*$ does not by itself determine the MZR.  Additionally,
the fact that the MZR and the $f_*$ plots look qualitatively different
from the nw and vzw models indicates that some process specific to the
cw model is affecting $f_*$ and $Z_g$ in similar ways.

What determines how $f_*$ scales in the cw model?  If $f_*$ depended
only on $\mlf$ then we would expect $f_* = 1/3$ at all scales; this
is indicated by the dashed line in the Figure.  The slow dependence
of $f_*$ on $M$ agrees qualitatively with the expected flat trend.
However, the normalization is much lower than expected, indicating that
another process must be suppressing galaxy growth.  Returning to energy
considerations, \citet{dek03} used the assumption that stars continue to
form until the total energy in outflows equals the virial energy of the
halo's baryons to predict that, in energy-driven wind scenarios such as
our cw model, $f_*$ (and hence the metallicity) should increase smoothly
with increasing mass below the reheating scale of $\approx150\;\kms$
and then flatten out above it.  The range of possible scalings that they
derive is indicated by the solid lines, where we have normalized their
scalings to $f_* = 1$ at the reheating scale.  Their predictions were
based on the assumption that energy in outflows couples efficiently
with the baryons in the galaxy's ISM and its halo.  It is clear from
Figure~\ref{fig:eta} that this key assumption does not hold in our
cw model.  At high masses the \citet{dek03} model predicts too little
suppression, indicating that outflows suppress star formation even if
much of the gas does not escape from the galaxy's halo.  Furthermore,
at low masses their model predicts too much suppression, indicating
that simulated outflows either escape these galaxies without entraining
the bulk of the halo gas (as would be expected if the outflows are not
spherically symmetric) and/or their energy is thermalized and radiated
away.  Thus, despite the impressive agreement between observations
and the simple model put forth by~\citet{dek03}, simulations
suggest that the interaction between outflows and ambient gas is
qualitatively different than what they assumed.  

We can speculate as to why good agreement with the MZR is obtained by
\citet{dek03}, as well as in semi-analytic models based on this type of
scenario such as~\citet{luc04}.  In these models, they assume that winds
from galaxies below the reheating scale inject progressively more energy
per baryon into ambient gas, thereby unbinding progressively more of
it to smaller masses.  In essence, they force a scaling of $\emlf$ with
$M_*$ that is similar to our vzw model (cf.~Figure~\ref{fig:eta_eff}),
by assuming that $\emlf\gg\mlf$ for small masses owing to efficient
energy coupling with ambient gas.  This physical process is not borne
out by our three-dimensional numerical simulations, in which reheating
is mostly irrelevant.  It is a minor coincidence that in constant wind
models, the blowout and reheating scales are fairly similar (equations
\ref{eqn:reheat} and \ref{eqn:blowout}).

In summary, a comparison between Figures~\ref{fig:mmr0} and~\ref{fig:eta}~
indicates that (1) in the presence of outflows, the scaling of $f_*$
(or, equivalently, gas fraction) depends more heavily on the mass-loading
factor $\mlf$ than on energetic considerations; and (2) $f_*$, while
a key driver, does not by itself determine the MZR.  Some other factor
must be important in determining the basic shape of the MZR.  We will
argue below that it is primarily gas accretion, although gas stripping
due to winds from neighboring galaxies can play a role in some situations,
as we will discuss in In \S~\ref{ssec:metret} and~\ref{ssec:suppress}.

\subsection{Metal Retention} \label{ssec:metret}
\begin{figure}
\setlength{\epsfxsize}{0.5\textwidth}
\centerline{\epsfbox{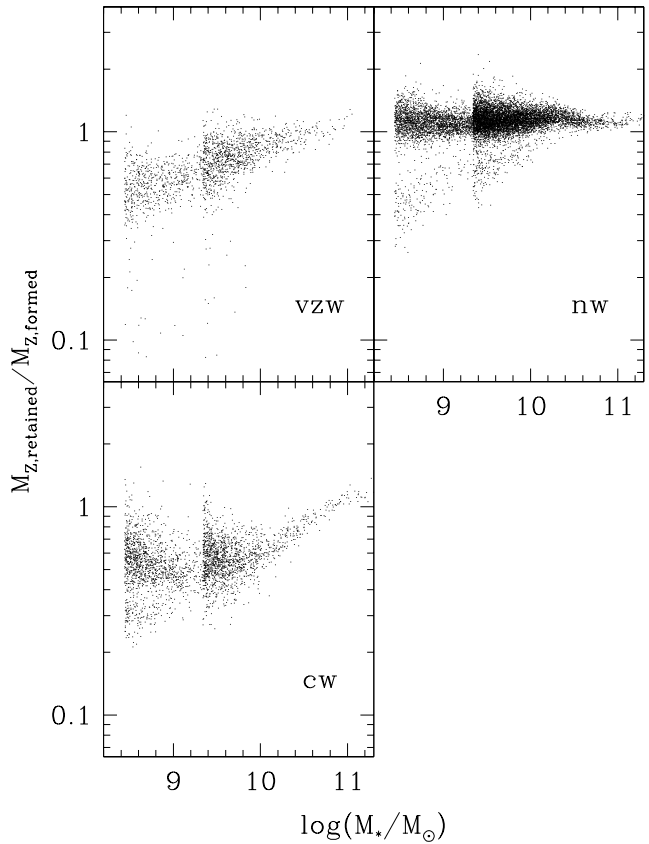}}
\vskip -0.0in
\caption{Fraction of metal mass retained $\fzret$ as a function of 
$\mstar$.  The two loci correspond to galaxies from the 16 and 
$32\hmpc$ volumes.  Without winds galaxies tend to retain their metals, 
whereas in the presence of winds galaxies can lose up to 50\% of their 
metals.  These trends are far too weak to account for the observed MZR.  
}
\label{fig:metret}
\end{figure}
The most popular interpretation of the observed MZR is that low-mass
galaxies exhibit low metallicity because they drive a larger fraction
of their metals into the IGM.  In order to determine whether this idea
explains the observable MZR in our simulations, we plot in 
Figure~\ref{fig:metret} the fraction of metals retained $\fzret$ as 
a function of stellar mass in each of our wind models.  If metal loss 
dominates the form of the observable MZR then we expect the slope and 
scatter of $\fzret$ as a function of $\mstar$ to mimic the MZR.  For
simplicity we discuss only the $16\hmpc$ boxes at $z=2$ here while 
noting that similar results hold for other scales and epochs.

In the nw model galaxies retain all of their metals on average (as
expected in the absence of outflows) although $\approx0.06$ dex of
scatter is introduced by dynamical disruption and uncertainties in
group idenfication.  This contrasts with the strong scaling ($Z_g
\propto \mstar^{0.28}$) and larger scatter (0.11 dex) in the nw MZR
(Figure~\ref{fig:mmr0}).  Moreover, a few low-mass galaxies 
have lost up to 80\% of their metals through interactions, yet we find
that even these galaxies exhibit no departure from the mean gas-phase
metallicities for their stellar masses.  Hence while they have ejected
significant baryons, those baryons were enriched at the same level as
the baryons that remained in the galaxies.

Turning to the wind models, at a fiducial stellar mass of $10^{10}\msun$
the cw and vzw models retain roughly 40\% and 70\% of their metals at
$z=2$, respectively.  However, the mean gas-phase metallicity of the 
cw galaxies is roughly 70\% {\it higher} than it is in the vzw galaxies.
Comparing the scalings reveals similar inconsistencies: In the vzw model 
$\fzret \propto \mstar^{0.07}$ 
while the MZR scales as $Z_g \propto \mstar^{0.21}$.
In the cw model, $\fzret \propto \mstar^{-0.04}$ with 0.09 dex of scatter, 
while its MZR scales as $Z_g \propto \mstar^{0.06}$ (note that this is 
in the opposite sense as the $\fzret$ scaling) with 0.19 dex of scatter.

The poor correspondence between the $\fzret$ scalings and the MZRs for
all three models indicates that galaxy metallicities are not primarily
driven by metal loss.  The inconsistency between the relative gas-phase
metallicities and $\fzret$ values of the vzw and cw models suggests
that the detailed way in which metals are distributed in different
baryonic phases must be taken into account.  Note that we do not claim
that $\fzret$ cannot trace the MZR \emph{in principle}; 
Figure~\ref{fig:metret} only shows that it does not do so in general.
Indeed, there is no rigorous reason why it should.  Observationally, it 
is possible to test this if the metallicities and masses in the stellar 
and gas phases as well as the net metal yield $y$ are known: if 
$Z_g \propto \fzret$ then it should also be proportional to 
$Z_*(y-M_g/M_*)^{-1}$ at all scales.  However, guided by our own 
simulations, we will argue in \S\ref{sec:compare_analytics} that the MZR
is dominated by the scaling of the star formation efficiency in the 
absence of winds, and by a competition between the rates of enrichment 
and dilution in the presence of winds.

\subsection{Suppression of $\mstar$ and $Z_g$} \label{ssec:suppress}
\begin{figure}
\setlength{\epsfxsize}{0.5\textwidth}
\centerline{\epsfbox{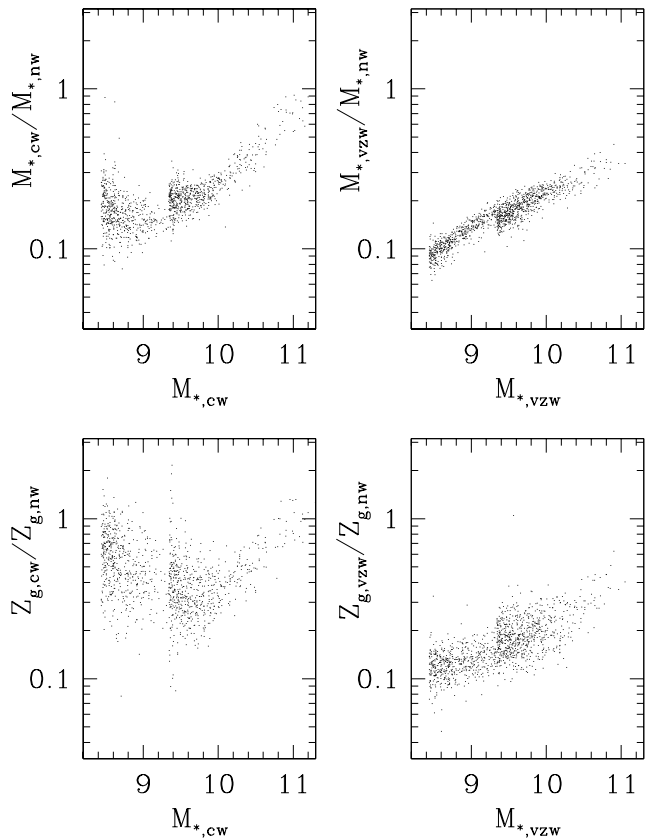}}
\vskip -0.0in
\caption{Ratio of stellar mass and gas-phase metallicity in the wind 
models versus the no-wind model at $z=2$.  The two loci correspond to 
galaxies from the 16 and $32\hmpc$ volumes.  In the cw model the effect 
of wind suppression does not vary strongly below the blowout scale whereas 
it decreases rapidly above it; in the vzw model wind suppression is less
effective at higher masses and the scaling is quite smooth.
}
\label{fig:matchGals}
\end{figure}

Another way to highlight the differences between the wind models is to 
compare their effects on individual galaxies.  Because all of our 
simulations were run with the same initial conditions, we can readily do 
this by matching the positions of individual galaxies between the 
simulations.  Figure~\ref{fig:matchGals} displays the ratios of stellar 
mass and metallicity in the wind models versus the no wind model at $z=2$.
Note that we have directly verified that excluding from 
Figure~\ref{fig:matchGals} those galaxies whose nw metallicities 
suggest $\emlf>0$ does not impact the slope, normalization, or scatter 
of the inferred suppression factors for either wind model.

In the cw model one might at first expect the ratios
$M_{*,\rm{cw}}/M_{*,\rm{nw}}$ and $Z_{g,\rm{cw}}/Z_{g,\rm{nw}}$ to be
$\approx 1/(1+\mlf)$ for galaxies below the blowout scale as long as
outflows couple inefficiently with the ambient gas (i.e.\ if $\mlf=2$ then
2/3 of the baryons should be ejected in winds).  Figure~\ref{fig:eta_eff}
showed that cw outflows are highly effective for all galaxies below the
blowout scale, in that their effective mass loading factor is, on average,
equal to $3/4$ of the true assumed $\mlf=2$, hence we might expect
$M_{*,\rm{cw}}/M_{*,\rm{nw}}$ to be roughly $1/(1+2.5) = 0.4$.  Turning to
Figure~\ref{fig:matchGals}, we see that $M_{*,\rm{cw}}/M_{*,\rm{nw}}$ lies
below this value even at the blowout scale, consistent with the existence
of an extra source of suppression implied by Figure~\ref{fig:eta}.
Above the blowout scale the cw ratios climb with tight scatter, reaching
unity at $\mstar \approx 10^{11.3}\msun$.

Below the blowout scale $M_{*,\rm{cw}}/M_{*,\rm{nw}}$ varies
slowly with decreasing mass and shows considerable scatter.  At the
lowest masses ($\mstar<10^9\msun$) $M_{*,\rm{cw}}/M_{*,\rm{nw}}$ and
$Z_{g,\rm{cw}}/Z_{g,\rm{nw}}$ actually climb with decreasing mass despite
the fact that $\emlf$ is constant in this range.  We have verified that
these trends continue to lower masses at higher mass resolution, hence
the behavior is not an artifact of numerical resolution.  Moreover, it
clearly conflicts with the na\"ive picture above.  In order to understand
it, we must consider how winds might affect galaxy growth in the full
context of structure formation rather than as isolated systems.

First, there is the possibility that an early generation of galaxies
pre-enriches the IGM, giving rise to a ``metallicity floor" in
the observable MZR.  Such a minimum has been inferred from quasar
absorption line systems~\citep[e.g.,][]{son01} as well as in our
simulations~\citep[e.g.][]{dav07}.  In \cite{dav06b} we hypothesized that
widespread pre-enrichment of the IGM in the cw model \citep{opp06}
may be responsible for its low-mass behavior.  One way to quantify
pre-enrichment is to measure how many metals were created in one galaxy,
expelled, and reaccreted onto another.  We measured the fraction of
galaxies' metals at $z=2$ that originated in another galaxy and find
that it is roughly 5\% and 15\% in the cw and vzw models, respectively.
While this is not insignificant, it is not sufficient to account for
the scatter in the cw MZR.  Furthermore, the fact that the vzw model
has more pre-enrichment but shows much less scatter suggests that
this process does not contribute significantly to the MZR scatter.
Hence the pre-enrichment hypothesis is unlikely to be correct.

Second, in a full hierarchical context there are two effects that reduce
the tendency of winds to drain galaxies' gas reservoirs, which may lead
to increased metallicities.  Whereas in a simple closed-box scenario
all gas is equally eligible to form stars, in a fully self-consistent
model galaxies' gas reservoirs possess density and pressure gradients and
star formation is concentrated in relatively small regions.  Gas that is
ejected from these areas can in principle be replaced on a dynamical time
by infall from less dense areas owing to the loss of pressure support.
A second effect is that gas entering a wind can potentially thermalize
its energy before escaping the galaxy's halo and fall back down onto the
galaxy in a galactic fountain, effectively increasing the galaxy's gas
accretion rate.  Figure~\ref{fig:eta_eff} does not indicate the presence
of scale-dependent galactic fountains below the blowout scale.  On the 
other hand, the fact that $M_{*,\rm{cw}}/M_{*,\rm{nw}}$ increases at 
low masses suggests that gas that escapes the galaxy in a wind is indeed 
being rapidly replaced, boosting $M_{*,\rm{cw}}/M_{*,\rm{nw}}$ above what 
would na\"ively be expected.  

Finally, galactic winds can affect neighboring galaxies.  In particular,
\citet{sca01} proposed that halos with masses $10^{9-10}\msun$ suffer
stripping of their baryons owing to winds from lower-mass neighbors.
In their model, once galaxies form they drive spherical winds whose
energies are comparable to the wind energies in our cw model.  The winds
from galaxies in low-mass halos ($<10^9\msun$) then strip baryons from
intermediate-mass halos ($10^{9-10}\msun$) that are not yet virialized
without affecting more tightly-bound massive halos ($>10^{10}\msun$),
with the result that galaxy formation in intermediate-mass halos is
suppressed (note that this effect is not related to tidal stripping).  
Although our winds are not spherical, when averaging over
a large sample of galaxies the effect should still be noticeable.  It 
is expected to be weak at $z\geq6$~\citep[e.g., Figure 7 of][]{dav06a} 
and to grow most noticeable during the heyday of galaxy formation 
$z\leq3$ \citep[Figure 8 of][]{sca01}.  Indeed, the ratios 
$M_{*,\rm{cw}}/M_{*,\rm{nw}}$ and $Z_{g,\rm{cw}}/Z_{g,\rm{nw}}$
show significantly less suppression at $z=6$ (not shown) than at $z=2$.
More interestingly, they also show less scatter at higher redshifts,
as expected for an effect whose strength depends on environment rather
than on galaxies' intrinsic properties.  In summary, it is likely that
stellar mass growth in the cw model divides into three mass regimes:
At low masses ($\mstar/\msun < 10^8$) galaxy growth is suppressed only
by outflows and hence $M_{*,\rm{cw}}/M_{*,\rm{nw}}\rightarrow 1/(1+\mlf)$. 
At intermediate masses ($10^8 < \mstar/\msun < 10^{10}$) galaxy growth is 
dominated by a collaboration between~\citet{sca01} suppression and 
losses to winds, where the relative contribution of each effect 
varies nontrivially with scale and epoch.  Because it only works in a 
particular mass regime,~\citet{sca01} suppression hence gives rise to a 
local minimum in $M_{*,\rm{cw}}/M_{*,\rm{nw}}$.  Finally, at high masses
($\mstar/\msun>10^{10}$) galaxy growth is again dominated by outflows, where
the effects of outflows weaken rapidly with increasing mass owing to the
rapidly declining $\emlf$ (Figure~\ref{fig:eta_eff}).

In the vzw model wind suppression of metallicity and stellar mass both
scale smoothly, with relatively little scatter, and without evidence
for a preferred scale.  We have already shown (Section~\ref{ssec:etastar})
that stellar masses in the vzw model can be attributed largely to 
the scaling of $\mlf$.  Given that $\fzret$ does not scale strongly with
mass in this model, it is also not surprising that 
$Z_{g,\rm{vzw}}/Z_{g,\rm{nw}}$ scales in the same way.

Does baryonic stripping occur in our vzw simulation? We can predict
whether it should be stronger or weaker in the vzw versus the
cw model by estimating the ratio of their momentum generation rates:
\begin{equation}
\frac{\dot{p}_{\rm vzw}}{\dot{p}_{\rm cw}} =
\frac{f_{*,{\rm vzw}}\eta_{\rm{W,vzw}}V_{\rm{W,vzw}}}
     {f_{*,{\rm cw}}\eta_{\rm{W,cw}}V_{\rm{W,cw}}}
\end{equation}
From Figure~\ref{fig:eta}, $f_{*,{\rm vzw}}\approx0.1$ and
$f_{*,{\rm cw}}\approx0.2$.  From before, we know 
$\eta_{\rm{W,vzw}} = 300\kms/\sigma$, $V_{\rm{W,vzw}} = 6.7\sigma$,
$\eta_{\rm{W,cw}} = 2$, and $V_{\rm{W,cw}} = 484\kms$, hence
we find that $\dot{P}_{\rm vzw}/\dot{P}_{\rm cw} \approx 1$---the
effect should be roughly as strong in the vzw model as in the cw 
model.  By subtracting the mean trend from the vzw MZR and 
inspecting the residual, we have found that vzw galaxies less
massive than $\mstar=10^{10}\msun$ do, in fact, exhibit signatures of
baryon stripping, seen as a change of slope and a slightly increased 
scatter in the MZR below this scale.  However, the effect is much 
less noticeable than in the cw model because it is small compared to
the effects of the strong outflows.  Applying this
idea to interpret observations, the tight scatter and smooth scaling in
the observed low-redshift MZR~\citep{lee06,tre04} indicate that either 
galactic winds do not carry enough momentum to suppress growth in 
neighboring halos appreciably or the effects of outflows must be
strong enough to erase any evidence of baryon stripping; both 
possibilities clearly conflict with the behavior of our cw model,
but are satisfied in our vzw case.

Comparing the top and bottom panels of Figure~\ref{fig:matchGals} yields
insight into the extent to which $f_*$ governs the MZR.  The metallicity
suppressions show considerably more scatter than the mass suppressions.
Additionally, the gas-phase metallicities are systematically about
50\% less suppressed than stellar masses in the cw models, independent
of $\mstar$.  (They are roughly equally suppressed in the vzw model.)
Both of these observations are inconsistent with the idea that $f_*$
solely governs the MZR; evidently the effects of winds on gas accretion
rates and star formation efficiencies also play a role.

In summary, Figure~\ref{fig:matchGals} indicates that outflows tend to
suppress both $\mstar$ and $Z_g$.  The trends and the levels of scatter
indicate that the amounts of suppression cannot, in general, be predicted
from simple scaling relations.  We have highlighted the fact that, in
both figures, the scatter in the cw model seems to be large below the
blowout scale and small above it, whereas the scatter in the vzw model
is comparable to $\sim0.1$ dex at all masses.  \citet{lee06} recently
reported that the $1\sigma$ scatter in the observed relation is roughly
0.1 dex from $10^6$--$10^{12}\msun$ at low redshift and argued that this
implied a less energetic form of metal-enhanced mass loss than blowouts.
Our results tend to support this view.  However, we also find that, in
order to avoid the large scatter introduced by~\citet{sca01} suppression,
winds must either transport significantly less momentum out of halos
than occurs in the simplest energy-driven wind models, or else they
must invoke large mass loading factors $\mlf>2$ at intermediate masses.
Momentum-driven winds naturally satisfies that requirement while our
constant wind model does not.

\section{Evolution of the MZR} \label{sec:metEvol}

In the previous sections we explored how the effects of winds scale with
mass by studying galaxies at a single epoch.  In order to understand
how outflows affect the \emph{evolution} of the MZR, we now discuss
how the metallicity evolves through cosmic time in our three models,
as a function of galaxy mass and gas fraction.  To do so, we trace the
evolution of each simulated galaxy at $z=2$ by searching for its most
massive progenitor in each simulation output, thereby compiling its mass
and metallicity history.  Because enrichment histories of individual
galaxies are highly stochastic, we bin the histories by stellar mass in
order to show the typical evolution as a function of stellar mass.

\subsection{Evolution of $M_*-Z_g$} \label{ssec:metEvol}
\begin{figure}
\setlength{\epsfxsize}{0.5\textwidth}
\centerline{\epsfbox{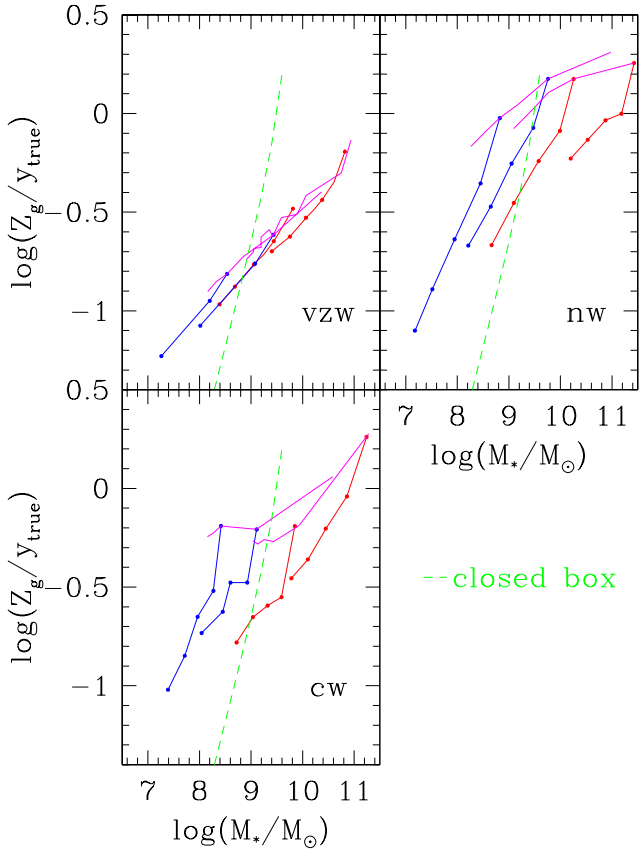}}
\vskip -0.0in
\caption{Mean enrichment histories of the simulated galaxies in the 
three wind models starting from $z=6\rightarrow2$.  Blue and red solid
lines denote evolutionary trends from the 16- and $32\hmpc$ boxes, 
respectively, while the green dashed line gives the slope of a closed-box 
with baryonic mass $5\times10^9 \msun$ ($Z_g/y=-\ln\mu(M_*)$).  
The magenta curves indicate the mean observable trends at $z=2$.  Galaxies 
evolve roughly parallel to the observable trend in the vzw model and more 
steeply in the cw and nw models.
}
\label{fig:mmr1}
\end{figure}

Figure~\ref{fig:mmr1} compares the mean enrichment histories from the
different models from $z=6\rightarrow2$.
As expected, galaxies generally increase in both mass and metallicity as
they evolve.  However, the slope of the evolution is in general neither
constant nor parallel to the observable trend owing to the fact that gas
accretion rates, star formation efficiencies, and wind properties vary
with scale and time.  In particular, galaxies do not generally evolve
as closed boxes (green dashed line) although the nw model comes quite
close even at high redshift.  Instead, their evolution is more shallow
owing to the fact that outflows expel gas that is enriched compared
to inflows from the IGM.  At lower redshifts where strong outflows
are rare and accretion rates are low, the evolution is expected to
more closely resemble the closed-box scenario.  This can be seen in
Figure 1 of~\citet{bro06}, where their simulated galaxies evolve from
$z=2\rightarrow0$ with a slope $\rm{d}\log(Z_g)/\rm{d}\log(M_*) \approx
1$.  Additionally,~\citet{sav05} have shown that closed-box scenarios
can account for the observed evolution from $z=0.7\rightarrow0.1$.

There is a slight upturn in the evolutionary slope from $z=3\rightarrow2$ 
that can be seen in all three models.  This feature occurs because gas 
accretion rates are declining during this interval owing to the increasing 
gas cooling times at intermediate overdensities; in other words, because
gas accretion grows decreasingly effective at diluting gas reservoirs, 
star formation grows increasingly effective at enriching them.  The fact
that the rate at which the normalization of the MZR changes during this
interval varies among our wind models illustrates how the time evolution
of the MZR, in addition to its normalization, shape, and scatter, is a
testable prediction of the wind model.

Turning to the individual models, galaxies in the nw model evolve nearly as 
steeply as a closed box in all but the highest mass bin.  The relatively 
steep evolution in the low-mass bins reflects the fact that these galaxies'
metallicities are not significantly diluted by inflows owing to their large
gas fractions.  By contrast, the shallower evolution in the highest mass bin
indicates that, owing to their high gas densities and star formation 
efficiencies, these galaxies possess somewhat lower gas fractions with
the result that inflows dilute their metallicities more readily.  We will
show in Figure~\ref{fig:zfg} that the same effects can be seen in these 
galaxies' effective yields.

In the vzw model galaxies enrich their gas reservoirs somewhat more
slowly.  This evolution can intuitively be understood as a consequence
of the tendency for vzw galaxies to expel a large fraction of their
metals.  However, we can also understand it via our analytical model
as follows: In the vzw model, the
effective mass loading factor scales with stellar mass as $\emlf\propto
\mstar^{-0.25}$ (Figure~\ref{fig:eta_eff}).  If galaxies' metallicities
remain near equilibrium (Section~\ref{ssec:analytic_imp}) and if the gas 
processing rate is approximately equal to the gas accretion rate 
$\sfr(1+\mlf)\approx\acc$, then the vzw MZR should scale with mass as 
$Z_g \approx y/(1+\mlf)$.  For $\mlf\gg 1$ (typical of small galaxies 
at high-$z$) we therefore expect $Z_g\propto \mstar^{0.25}$---exactly 
the scaling seen in Figure~\ref{fig:mmr1}.

The evolution of the MZR in the cw model is more complex than in the other
two models owing to the nontrivial environment-dependent interactions
between the outflowing winds and the inflowing gas.  The relevant points
are that the basic shape of the constant wind MZR does not evolve with
redshift while its normalization increases somewhat more rapidly than
that of the vzw but not as rapidly as the nw model.

\subsection{Evolution of Effective Yields} \label{ssec:yeff}
\begin{figure}
\setlength{\epsfxsize}{0.5\textwidth}
\centerline{\epsfbox{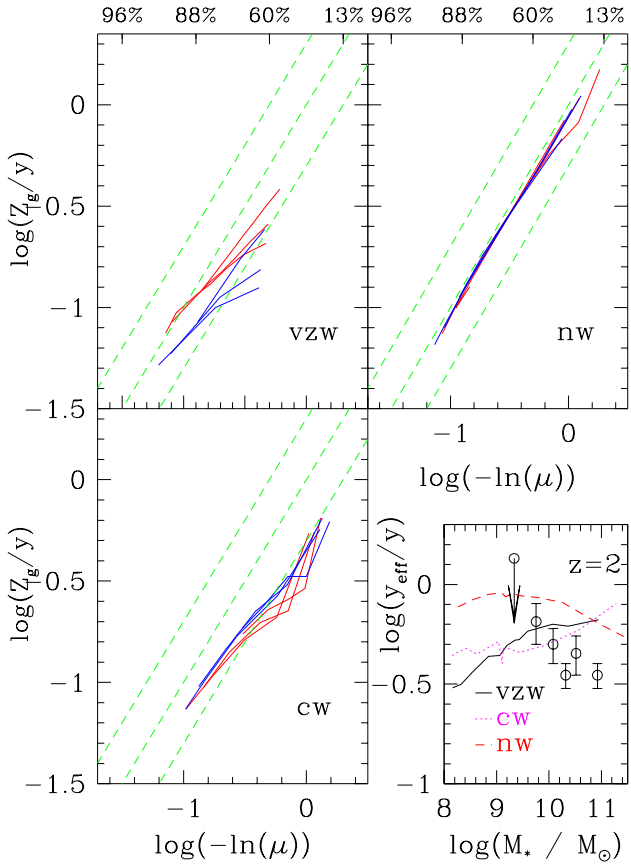}}
\vskip -0.0in
\caption{Mean enrichment histories of the simulated galaxies in the 
three wind models starting from $z=6\rightarrow2$.  Dashed green lines
show where galaxies fall that have $y_{\rm{eff}}/y=0.5,1.0,2.0$. Blue and 
red solid lines denote evolutionary trends from the 16- and $32\hmpc$ 
boxes, respectively.  The nw galaxies involve much like closed boxes 
while the vzw and cw galaxies show suppressed effective yields.
}
\label{fig:zfg}
\end{figure}

A more direct way to investigate the extent to which our simulated
galaxies depart from closed-box evolution is to plot the metallicity
versus gas fraction, Figure~\ref{fig:zfg}.  The axes in this
figure are chosen so that galaxies with a constant effective yield
$y_{\rm{eff}}\equiv Z_g/-\ln(\mu)$ (where $\mu$ denotes the mass 
fraction of baryons in the gas phase) evolve along straight lines.
The dashed green lines show the evolution for (from bottom to top)
$y_{\rm{eff}}/y=0.5,1.0,2.0$.  Individual galaxies generally evolve from
lower left to upper right in this space.

The nw galaxies remain quite close to the closed-box curve
$y_{\rm{eff}}/y=1.0$ as suggested by the steep evolution
in Figure~\ref{fig:mmr1}.  The effective yield never exceeds the
closed-box case, as required by Theorem 3 of~\citet{edm90}.  It drops
below the closed-box value most strongly at the redshift corresponding
to the peak accretion rates $z\approx 3$ because the dilution of the
metallicities at this epoch is overcompensating for the increase in gas
fraction (this happens irrespective of the wind model and is therefore a
robustly predicted---if difficult to confirm---consequence of the global
gas accretion peak at $z\approx3$).  After the accretion rates begin
subsiding, $y_{\rm{eff}}/y$ returns to the closed-box value relatively
quickly owing to continued star formation~\citep{dal06}.

By contrast, galaxies in the vzw model tend to evolve from higher to
lower $y_{\rm{eff}}/y$ during this period owing to the combined effects
of dilution and outflows.  In this model accretion rates fall off
somewhat more slowly following the $z=3$ peak than in the other models,
pushing the peak of star formation to lower redshift~\citep{opp06} and
delaying the recovery of the effective yields.  The delayed falloff in
accretion owes to galactic fountain effects that occur because outflowing
gas does not escape the halo $\sim20-50\%$ of the time, on average.
The gas is then retained in a puffy distribution, owing to small wind
speeds that do not drive gas far from the galaxy, and becomes available
for reaccretion on a time scale smaller than a Hubble time.

Galaxies in the cw model show a gradual evolution from nearly closed-box
yields $y_{\rm{eff}}/y \approx 0.8$ at $z=6$ to lower values 
$y_{\rm{eff}}/y \approx 0.3$ at $z=3$ before rebounding to 
$y_{\rm{eff}}/y \approx 0.5$ at $z=2$.  The relatively weak dependence
of $y_{\rm{eff}}/y$ on $M_*$ at all redshifts recalls the flat MZR in
Figure~\ref{fig:mmr0} and owes to the fact that most of the galaxies in 
this figure lie below the blowout scale and are hence roughly equally 
affected by winds.

The bottom-right plot in Figure~\ref{fig:zfg} compares how $y_{\rm{eff}}$
varies with $M_*$ at $z=2$ in our three models versus observations
of UV-selected galaxies at $z\sim2$~\citep{erb06}.  The nw model
shows no trend at low masses and declines with increasing mass for
masses above $10^{10}\msun$.  The decline occurs because gas fractions
decline with increasing mass: Galaxies more massive than $10^{10}\msun$
have gas fractions $\mu\leq20\%$ in the nw model so that unenriched
inflows are able reduce $y_{\rm{eff}}$ efficiently~\citep{dal06}.  The 
$y_{\rm{eff}}$ behavior is qualitatively consistent with the observed 
trend although the normalization is $\approx50\%$ too high.

In the vzw model, $y_{\rm{eff}}$ increases with increasing mass at low
masses and flattens out around $10^{10}\msun$.  This behavior at face
value conflicts with observations at $z\sim2$.  On the other hand,
it bears a striking resemblance to the observed trend at low redshift
\citep{gar02,tre04,pil04}.  Although we do not show it here, the overall
shape of the $y_{\rm{eff}}$ trend in the vzw model does not vary with
redshift.  An aggressive interpretation of Figure~\ref{fig:zfg} would be
that winds at high redshifts must differ qualitatively from winds at low
redshifts, with the former relatively more effective in massive galaxies
than the latter; in other words, at high redshift the effect of winds on
$y_{\rm{eff}}$ should preserve the nw scaling while at low redshift the
vzw model is more realistic.  However, considering that~\citet{erb06}
were forced to infer gas masses indirectly rather than measuring them,
and given that their measurements span a much smaller dynamic range than
low-redshift observations, we prefer not to draw any firm conclusions from
Figure~\ref{fig:zfg}.  Future measurements that trace high-redshift gas
masses more directly (e.g. with ALMA) will constrain the effective
yields of high-redshift galaxies more directly, and will provide a 
key test of the momentum-driven wind scalings.

The vzw trend at low masses is expected despite their higher gas 
fractions (not shown) because $\mlf$ increases to lower masses 
(Figure~\ref{fig:eta_eff}).  In terms of our analytical model, we now 
anticipate the conclusions of Section~\ref{sec:compare_analytics} 
by assuming that $\sfr(1+\mlf) = \acc$ and dividing 
Equation~\ref{eqn:Zg_eq} by $y\ln(1/\mu)$ to find the 
equilibrium condition:
\begin{equation}
\frac{y_{\rm{eff,eq}}}{y} = \frac{1}{(1+\mlf)\ln(\frac{1}{\mu})}
\end{equation}
As mass increases, $\mlf$ decreases (Figure~\ref{fig:eta_eff}), 
hence $y_{\rm{eff,eq}}$ increases.

The flattening behavior in $y_{\rm{eff,eq}}$ around $\mstar=10^{9.5-10}\msun$ is more
interesting.  Recall that in our vzw model the fraction of baryons
lost to winds does not vary strongly with mass (Section~\ref{ssec:metret}).
Evidently, a flattening in $y_{\rm{eff}}$ does not necessarily indicate
a scale at which superwind feedback becomes effective at removing a
galaxy's metals, an interpretation we discuss further below.  Nor does 
it indicate that $y_{\rm{eff}}$ has reached the true yield $y_{\rm{eff}}/y=1$.  
Instead, it seems to be a coincidence resulting from a competition between 
dilution owing to accretion and the outflows' suppression of star formation.
Indeed, in more massive galaxies ($M_* >10^{11}\msun$) $y_{\rm{eff}}$
begins to increase with mass again owing presumably to the fact that
gas accretion rates are falling in this range.  While the highest mass
bin in the~\citet{tre04} data tantalizingly suggests agreement with this
behavior, confirmation will have to await the arrival of larger samples
of massive star-forming galaxies at $z\sim1$ (since at $z=0$ massive galaxies
are generally too gas-poor for their gas-phase abundances to be measured).

\citet{gar02} also found a flattening of $y_{\rm{eff}}$ below
$\sim100\;\kms$ at low redshift, but his interpretation was that
low-mass galaxies retain a smaller fraction of their metals owing
to the onset of winds in less massive galaxies, in conflict with our
interpretation.  However, \citet{dal06} recently used the same data to
show that the observed trend in $y_{\rm{eff}}$ can be obtained without
assuming that winds are more effective in low-mass galaxies by removing
the approximation that the gas fraction is constant.  In this view,
$y_{\rm{eff}}$ is suppressed in low-mass galaxies because their low
star-formation efficiencies prevent their $y_{\rm{eff}}$ from recovering
quickly from inflow episodes.  Our findings tend to support the latter
view.

In the cw model, $y_{\rm{eff}}$ does not vary with mass below the blowout
scale and increases with increasing mass above it, emphasizing the idea that
the effectiveness of winds below the blowout scale does not seem to follow
the trends that would be expected from simple scaling arguments.

Effective yields can also give insight into another important question
in the study of high-redshift galaxies, namely whether the buildup of
stellar mass occurs in a predominantly smooth or episodic fashion.
At redshifts $z<1$, the tight observed correlation between stellar
mass and $\sfr$ argues in favor of a predominantly smooth mechanism for
relatively massive galaxies $10^{10}<\mstar/\msun<10^{11}$~\citep{noe07a}.
At $z\sim2$, clustering measurements indicate that Lyman-break and
submillimeter galaxies possess duty cycles of $\sim1$~\citep{ade05a}
and $\sim0.1$~\citep{bou05}, respectively, suggesting that Lyman 
Break galaxies form stars smoothly while submillimeter galaxies are
more bursty.  Similarly, recent clustering measurements indicate
that Lyman-$\alpha$ emitters at $z\approx4.5$ possess duty cycles
of $\sim10$\%, again hinting at relatively bursty star formation
histories~\citep{kov07}.

Effective yields provide another way to test this behavior, because in
the absence of inflows and outflows, galaxies' effective yields quickly
recover to the true yield.  Hence if the timescale for the $y_{\rm{eff}}/y$
to recover to the true yield is short compared to the duty cycle then some mechanism must be actively
suppressing it~\citep{dal06}.  Quantitatively, if we define the ratio
of the effective yield to the true yield $X_y \equiv y_{\rm{eff}}/y$
then, in the absence of inflows and outflows, $X_y$ varies with time
according to \begin{equation} \label{eqn:dXydt} \frac{{\rm d}\;X_y}{{\rm
d}\;t} = \frac{\sfr}{M_g \ln(\frac{1}{\mu})}(1 - X_y) \end{equation}
This equation shows that the equilibrium solution $X_y = 1$ is a stable
one~\citep{kop99} and that departures from equilibrium disappear with an
e-folding timescale given by $M_g\ln(1/\mu)/\sfr$.  Applying this timescale 
to UV-selected galaxies at $z\sim2$, where by computing weighted means
over bins 2--6 in~\citealt{erb06} we determine $(<\sfr>,<Mg>,<\mu>)
\approx (29\;\msun\rm{yr}^{-1}, 2.1\times10^{10}\;\msun, 0.38)$, we find 
that the timescale for $y_{\rm{eff}}/y$ to return to the closed-box yield 
is 700 Myr.  This is considerably shorter than the gas consumption time 
of 1.2 Gyr.  This short timescale together with the suppressed observed 
effective yields $y_{\rm{eff}} < 0.01$ imply that star formation in these
galaxies cannot be episodic in nature, consistent with the large inferred
duty cycles.  In our simulations, this timescale is less than 50\%
of the Hubble time at all epochs in both wind models, indicating that
our predicted effective yields reflect star formation and gas accretion
processes that are predominantly smooth rather than episodic in nature.

In summary, at high redshift ($z=6\rightarrow2$) galaxies tend to evolve
from high to low $y_{\rm{eff}}$ as accretion rates increase and are hence
expected to ``rebound" at lower redshifts $z<2$ where gas accretion rates
decrease.  Superwind feedback generically suppresses $y_{\rm{eff}}$ in our
galaxies as expected.  The resulting $M_* - y_{\rm{eff}}$ trend depends on
the ways in which winds affect how inflows, outflows, and gas fractions
scale with mass.  The cw trend conflicts with high- and low-redshift
observations while the vzw trend agrees qualitatively with low- but not
all high-redshift observations.  The fact that the vzw model produces
better agreement with the directly-measured high-$z$ MZR and low-$z$ $M_*
- y_{\rm{eff}}$ trends suggests that our vzw model is closer to reality.
In both cases star formation occurs via predominantly smooth rather than
bursty processes.  It will be interesting to see whether the vzw model's 
conflict with the observed trend in $y_{\rm{eff}}$ at high-$z$ is 
alleviated once direct measurements of gas densities at high redshifts 
become available.

\subsection{Analytic Model MZR Evolution} \label{ssec:compare_evol}
\begin{figure}
\setlength{\epsfxsize}{0.5\textwidth}
\centerline{\epsfbox{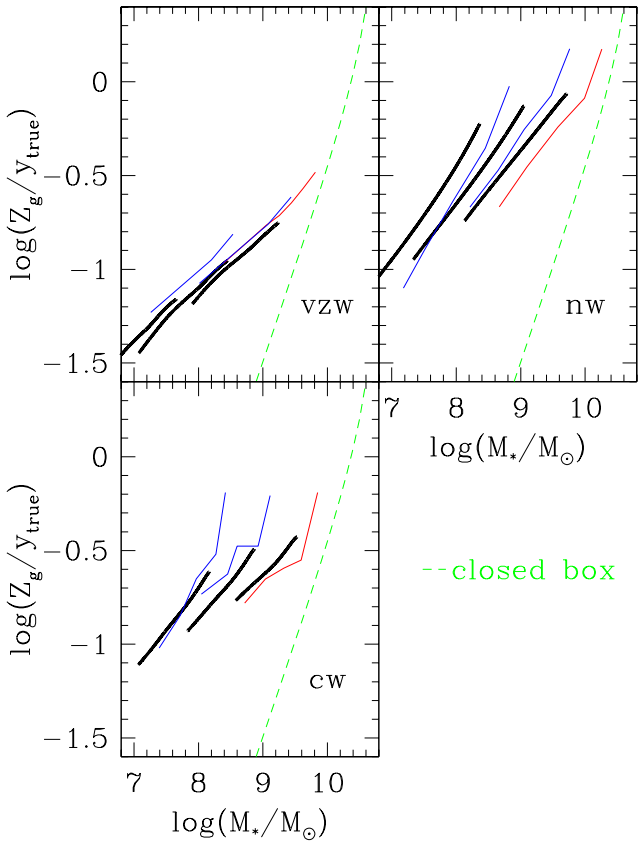}}
\vskip -0.0in
\caption{A comparison of the mean evolution of galaxies through the MZR in
the w16 (thin blue) and w32 (thin red) models versus our analytical model.
The analytical model reproduces the qualitative differences between the wind 
models although it does not reproduce the behavior of the individual models
in detail.  In particular, the analytical model yields a nonzero slope at low
masses in the cw model, in clear conflict with the hydrodynamic simulations.
}
\label{fig:hydro_analytic}
\end{figure}

In Section~\ref{sec:compare_analytics} we will synthesize the insights
gained from our analytical model to understand the origin of the MZR in
our simulations.  In order to justify this analogy, in this section we
show that our analytical model broadly reproduces the evolution of the 
simulated MZR.

Figure~\ref{fig:hydro_analytic} compares how galaxies evolve through the
MZR in our simulations (thin lines) versus our analytical model (thick
lines) for three different mass bins.  The analytical model succeeds in
recovering most of the qualitative features of the models as well as the
differences between them.  For example, in each model the observable
MZR has nonzero slope.  This is only achieved if the star formation
efficiency or the $\mlf$ scales with mass (Section~\ref{ssec:ns_nowind}).
The agreement between the simulated and analytical trends results
because we directly tuned our analytical star formation efficiencies to
match Figure~\ref{fig:simfits}.  We have verified that our analytic
model reproduces the simulated galaxies' effective yields as well
(Figure~\ref{fig:zfg}) although we do not discuss it here.

The qualitative differences in evolution between our different 
outflow models are well-reproduced.  For
example, nw galaxies evolve the most from $z=6\rightarrow2$ while cw 
galaxies evolve the least because the nw galaxies retain their entire 
gas reservoirs while the accretion histories in the cw model peak at an
earlier time.  The vzw galaxies evolve the most shallowly owing to their
high $\mlf$'s while the nw galaxies evolve the most steeply.  The result
of these differences is that the vzw galaxies are the least enriched at
$z=2$ while the nw galaxies are the most enriched.

While our analytical model reproduces many of the gross features of the
simulations as well as the differences between the different wind models,
it does a poorer job of reproducing the individual wind models in detail.
For example, in the nw and vzw models metallicities at $z=2$ are 
$\approx0.1$ dex too low.  Such offsets can easily result from minor 
inconsistencies in accretion histories or the definition of the gas-phase 
metallicity.  The upturn at late times is not 
conspicuous in any of our analytical models although it can readily be 
reproduced by forcing gas accretion rates to drop more precipitously at 
$z<3$ than we have done.  This is especially clear in the cw model, which
is generally the most difficult model to reproduce owing to the fact
that its behavior does not scale smoothly with galaxy mass.  Our analytical 
cw model inevitably yields a nonzero slope in the observable MZR, in 
conflict with the simulations.  This can be alleviated by allowing the
star formation efficiencies $\sfr/M_g$ to vary more slowly at low masses
than at high masses (Figure~\ref{fig:nw_comp_sfr}).  However, further 
fine-tuning of the star-formation efficiencies and accretion histories in 
our analytical model would yield little insight.  The important point to 
take away from Figure~\ref{fig:hydro_analytic} is that our analytical model 
captures the essential ingredients that determine how galaxies evolve 
through the MZR in our fully three-dimensional simulations.  In the next
Sections, we will therefore apply our analytical model to determine the
conditions that drive the form of the observable MZR.

\section{Understanding the Mass-Metallicity Relation} \label{sec:compare_analytics}

In this section we use the intuition gained in the past several
sections along with the simulation results in order to piece together a
comprehensive understanding of what drives the MZR's form and evolution.
We have already demonstrated that our vzw simulation produces good
agreement with the slope, normalization, and scatter of the observed
MZR (Figure~\ref{fig:mmr0}) and that our analytical model provides an
acceptable description of how galaxies evolve in our full simulation
(Figure~\ref{fig:hydro_analytic}).  Hence we begin by showing how our
analytical model can account for the amplitude, slope, and scatter of the
simulated---and, by implication, the observed---MZR.

\subsection{Implications of the Model} \label{ssec:analytic_imp}

\subsubsection{Normalization and Scaling} \label{sssec:norm}

Combining our original analytical model, equation~\ref{eqn:evol_Mz},
and the evolution of the gas mass, equation~\ref{eqn:evol_Mg}, it is
straightforward to show that galaxies evolve through the MZR with a
slope given by
\begin{equation}
\frac{{\rm d}\;Z_g}{{\rm d}\;\mstar} = \frac{1}{M_g} \left(
\frac{\acc}{\sfr}Z_g(\alpha_Z - 1) + y \right) \label{eqn:dZgdMs},
\end{equation}
which is equal to zero if
\begin{equation}
Z_g = y \frac{\sfr}{\acc(1-\alpha_Z)} \equiv \zeq \label{eqn:Zg_eq}
\end{equation}
This possible balance between the influences of star formation and 
infall has been identified previously~\citep[e.g.,][]{tl78,kop99}.
Winds enter into the determination of $\zeq$ indirectly by 
modulating the rate at which a galaxy depletes its gas reservoir 
as well as the relative enrichment of the satellite galaxies that 
it accretes.  If a galaxy processes its gas into stars and winds at 
the gas accretion rate\footnote{This can be viewed as setting the 
constant $k\equiv\sfr/\acc$ in~\citet{tl78}.}, 
then $\sfr(1+\mlf) = \acc$, which in turn yields the equilibrium 
gas-phase metallicity $\zeq = y(1+\mlf)^{-1}(1-\alpha_Z)^{-1}$.  
In what follows we will show that in our wind models this is 
approximately true.

\subsubsection{Scatter} \label{sssec:scatter}
The ratio of a galaxy's metallicity to its equilibrium metallicity 
$X_Z\equiv Z_g/\zeq$ evolves with time according to
\begin{equation}
\frac{{\rm d}\;X_z}{{\rm d}\;t} = \frac{\acc}{M_g}(1 - X_Z) - yX_Z
\left(\frac{\ddot{M}_{\rm SFR}}{\sfr} - \frac{\ddot{M}_{\rm ACC}}{\acc}\right) \label{eqn:dXzdt}.
\end{equation}
The second term in Equation~\ref{eqn:dXzdt} is small except during 
short-lived interactions, hence we may neglect it.  In this case,  
Equation~\ref{eqn:dXzdt} implies that the equilibrium solution $X_Z=1$ 
is a stable one~\citep{kop99} and that departures from equilibrium 
disappear on a timescale given by $M_g/\acc$, or the timescale for the 
gas reservoir to be diluted by a factor of 2.  If this timescale is 
shorter than the timescale over which perturbations to a galaxy's 
metallicity occur then galaxies' gas-phase metallicities recover from 
perturbations quickly, suppressing scatter in the observable MZR.

\subsection{Normalization and Scaling Without Outflows} \label{ssec:ns_nowind}
\begin{figure}
\setlength{\epsfxsize}{0.5\textwidth}
\centerline{\epsfbox{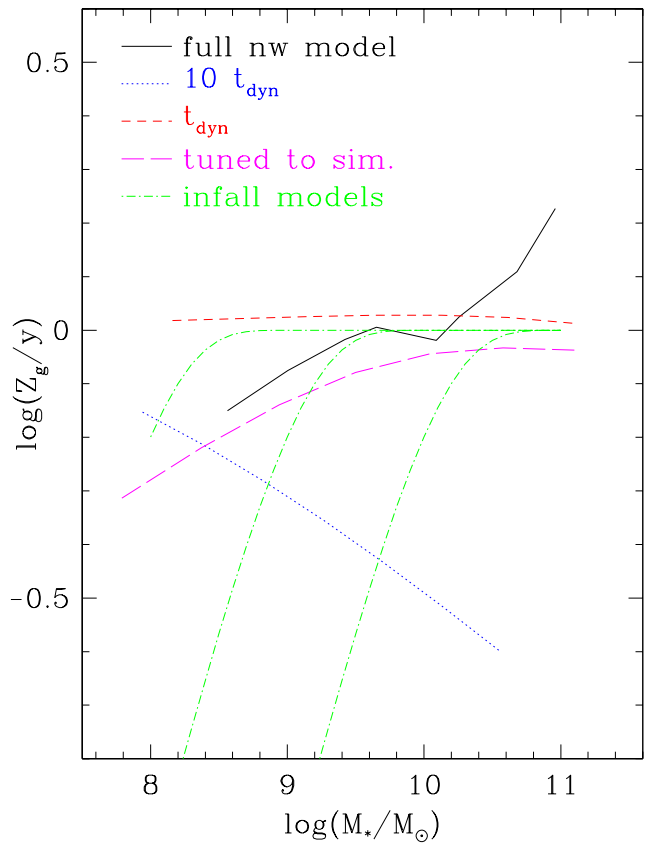}}
\vskip -0.0in
\caption{Predicted MZR at $z=2$ in the nw model versus several 
analytical models.  The solid black curve gives the mass-weighted 
gas-phase metallicity in the nw simulation.  The red dashed and blue 
dotted lines are computed by assuming that gas collapses into stars in 1
and 10 dynamical times, respectively (see text); the long-dashed magenta 
line is computed by tuning star formation efficiencies to match the nw 
model.  Green dot-dashed curves denote infall models (Equation~\ref{eqn:infall}) 
assuming constant gas masses of $10^8$, $10^9$, and $10^{10}\msun$ from
left to right.  The slope of the MZR in the absence of winds is dominated by
the scaling of the star formation efficiency.
}
\label{fig:nw_comp_sfr}
\end{figure}

We begin our discussion of the origin of the MZR with the no-wind 
scenario in order to develop some intuition about how hierarchical 
structure formation impacts the MZR.  In Figure~\ref{fig:mmr0} we showed 
that its MZR has approximately the correct slope, but an amplitude that 
is $\approx0.5$~dex too high.  We further showed in 
Figure~\ref{fig:eta_eff} that no-wind galaxies retain most of their gas, 
as expected without outflows.  Hence the only physical effect left that 
can cause a slope in the MZR is the star formation efficiency.

If galaxies converted their gas into stars at precisely the gas
accretion rate then the slope of the MZR in the nw model would be
zero.  This can be seen from the ``infall model" formalism 
of~\citet{lar72}, which tells us that if the gas mass is constant and
$\sfr=\acc$ then the gas metallicities evolve as:
\begin{eqnarray} \label{eqn:infall}
Z_g & = & y (1-e^{-\nu})\\
\nu & \equiv & \mu^{-1} - 1, \nonumber
\end{eqnarray} 
(Note that this can be obtained from Equation~\ref{eqn:dZgdMs} by 
substituting $\sfr/\acc=1$.)  In this model, for sufficiently small gas 
masses the MZR would be flat because the gas fraction $\mu$ would rapidly 
shrink to zero for all galaxies.  Equivalently, Equation~\ref{eqn:dXzdt}
tells us that we would expect all galaxies to approach $\zeq=y$ as long 
as $M_g/\acc$ were significantly less than the hubble time.  The MZR 
clearly is not flat in the nw model.  Therefore, guided by our discussion 
of Figure~\ref{fig:eta}, we now ask whether the scaling of the star 
formation efficiency $\sfr/M_g$ can account for the scaling of the 
no-wind MZR.

We have calculated the $z=2$ MZR using our analytical model with three
different prescriptions for the SFR.  In the first, we compute
the star formation rate by assuming that gas condenses into stars in
10 dynamical times; this is similar to the~\citet{ken98} relation.
The gas densities are obtained from the baryonic masses by combining 
the Virial Theorem with the low-redshift baryonic Tully-Fisher 
relation~\citep{geh06} and assuming that galaxies are 5 times as dense 
as their host halos (the value chosen for this ratio affects the amplitude
but not the trend of our results).  In the second, we assume that gas 
condenses into stars in one dynamical time.  In the third, we tune the 
star formation efficiencies $\sfr/M_g$ to reproduce the efficiencies 
in the nw simulations.

Figure~\ref{fig:nw_comp_sfr} compares the resulting MZRs at $z=2$ with
the trend from the fully three-dimensional nw model using mass-weighted
gas-phase metallicities (solid black curve).  Additionally, we have 
plotted the evolution of $Z_g$ versus $M_*$ for three representative 
infall models (Equation~\ref{eqn:infall}; green dot-dashed curves)
corresponding to constant gas masses of $10^8, 10^9$, and $10^{10} \msun$
from left to right.  The trend from the 10-dynamical time prescription has 
a negative slope, indicating that more massive galaxies possess larger gas 
masses (and larger gas fractions) than less massive galaxies and hence cannot 
enrich their gas reservoirs as effectively; in terms of 
Equation~\ref{eqn:dZgdMs}, since $\sfr/\acc$ declines with increasing $M_*$,
$Z_g$ does as well.  The trend from the one dynamical time model is flat, 
as expected for a scenario with such efficient star formation; indeed, for
this model $\sfr\approx\acc$ so that Equation~\ref{eqn:infall} accurately
predicts $Z_g/y=1$ for all masses since the gas fractions are
negligible.  By contrast, the trend from the model in which we have tuned 
the star formation efficiencies exhibits the desired positive slope.  This
trend is in qualitative agreement with the numerical trend although the 
most massive nw galaxies exhibit metallicities above the yield.  The high 
metallicities at the massive end are likely a consequence of accreting 
pre-enriched gas; in terms of Equation~\ref{eqn:dZgdMs}, $\alpha_Z>0$.

The simulated trend is shallower than expected from Equation~\ref{eqn:infall} 
because the assumption $\sfr=\acc$ is violated weakly in the 
absence of outflows; in particular, $\sfr/\acc$ drops from
1.0 at $10^8\msun$ to 0.9 at $10^{11}\msun$ in the analytical (magenta) 
curve while $\mu$ drops from 0.6 to 0.2 over the same interval.  
In short, the reason the nw model has the correct MZR slope is that 
the star formation efficiency in a hierarchical structure 
formation scenario naturally yields the desired differential with galaxy 
mass (cf.\ \S~\ref{ssec:etastar}).  Nevertheless, the excessive amplitude
(cf.\ Figure~\ref{fig:mmr0}) suggests that metals must be preferentially 
removed from these galaxies.  Hence outflows are necessary
to obtain the correct amplitude~\citep{kob07}, but are {\it not} required 
to obtain the suppression of metallicity in low-mass 
galaxies~\citep{ros06,tas06}.

\subsection{Normalization and Scaling With Outflows} \label{ssec:ns_wind}

Next we consider the impact of winds.  From Figure~\ref{fig:matchGals}
we see that outflows suppress both the stellar mass and the metallicity.
In order to lower the MZR amplitude, outflows must lower $Z_g$ more
than $\mstar^{0.3}$, which is the observed slope of the MZR.  

Looking first at $\mstar=10^{10}M_\odot$, which is around $L_*$ at $z=2$, 
we see from Figure~\ref{fig:mmr0} that both the cw and vzw models produce 
roughly the correct MZR amplitude (recall that the observed metallicities
and the yield each introduce $\approx0.3$ dex of uncertainty).  
Under the assumption that the MZR is governed primarily by the mass 
loading factor $\mlf$ via the equilibrium conditions 
(cf.\ \S~\ref{sssec:norm}), this is expected because galaxies 
in both models experience effective mass loading factors in the range
$\emlf = 1-1.5$ at this scale (cf.\ Figure~\ref{fig:eta_eff}).
Intuitively, $\emlf$ determines the level of suppression of gas 
enrichment and stellar mass growth because higher mass-loading
factors lead to lower gas densities (cf.\ Figure~\ref{fig:simfits}) 
and to more gas being ejected in winds rather rather than converted into 
stars.  This leads to a tendency for galaxies with similar $\emlf$ 
to possess similar properties regardless of wind model.

Above this mass scale, constant wind outflows cannot escape halos,
causing $\emlf$ to drop rapidly.  Hence cw metallicites grow much more
rapidly with $M_*$ than observed.  Below this scale, metallicities reach
a minimum in the range $\mstar = 10^{9.5}\msun$ and then ``rebound"
slightly to lower masses.  If this trend continues to $z=0$ (as it does
from $z\approx 6\rightarrow 2$), then this would be in gross conflict
with observations.  The cause of this scaling likely owes to the flat
trend of $\emlf$ with mass at low masses, with the slight rebound at
low masses due to baryon stripping from nearby galaxies' outflows as
discussed before (see also~\citealt{sca01}, Figure 12).

In the vzw case, galaxies are always below the blowout scale (which
increases linearly with mass).  In this regime, the mass loading factor
governs the MZR.  In order for the scaling of $Z_g\propto M_*^{0.3}$
to hold, $\mlf$ must be roughly proportional to $M_*^{-0.3}$, which is
satisfied in the vzw case.  There is also slight evidence for baryon
stripping, but it is highly subdominant compared to the high $\emlf$
in the mass range where stripping is effective.

In summary, winds suppress galaxy masses and metallicities primarily by
modulating the relative rates at which gas reservoirs are enriched and
diluted and secondarily by stripping baryons from neighboring galaxies.
In the vzw model the latter effect is small compared to the former
while in the cw model both are significant.  We emphasize that the
normalization and scaling of the MZR are {\it not} determined by the
total fraction of metals that galaxies retain because the rates of gas
accretion and star formation are too rapid for the gas reservoirs to
retain any memory of this quantity.  If baryon stripping is negligible
compared to the effects of outflows then the most important parameter
is the effective mass loading factor $\emlf$; if the wind speed exceeds
the escape velocity then $\emlf\propto\mlf$ and the scaling of $\mlf$
dominates the MZR.  

\subsection{Normalization and Scaling: The Equilibrium Metallicity} \label{ssec:ns_zeq}
\begin{figure}
\setlength{\epsfxsize}{0.5\textwidth}
\centerline{\epsfbox{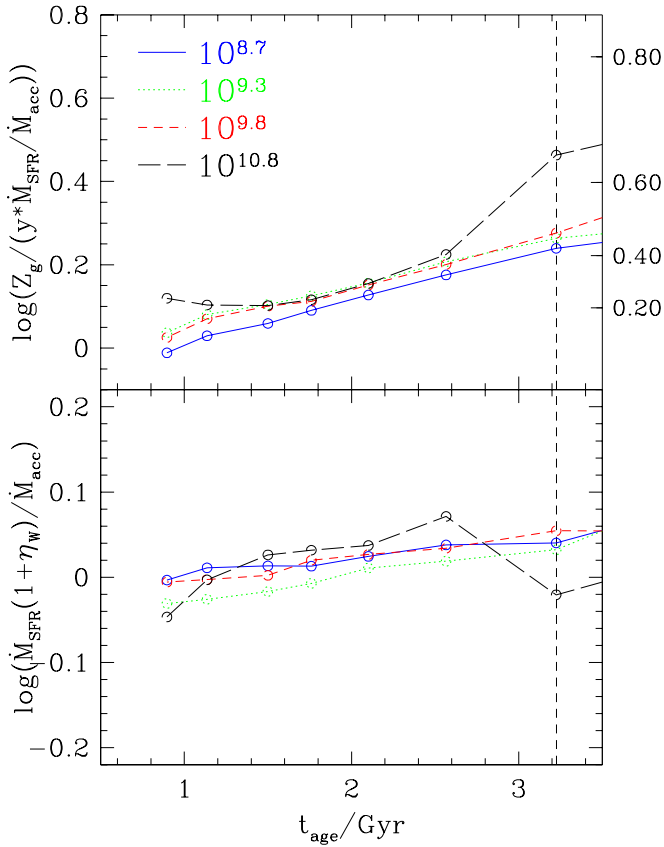}}
\vskip -0.0in
\caption{(Top) Mean ratio of actual to equilibrium gas-phase metallicity 
in four bins of $\mstar$ in the vzw model.  The different curves 
correspond to galaxies in different mass bins; the legend gives the 
stellar masses at $z=2$.  The vertical dashed line denotes $z=2$.  
(Bottom) Mean ratio of the gas processing to gas 
accretion rates.  The mean metallicity of infalling material grows 
with both galaxy mass and time and is roughly $\alpha_Z=50\%$ of the mean 
ISM metallicity at $z=2$ (right y-axis on top panel).  Gas processing 
rates generally lie within 20\% of the equilibrium values at all times.
}
\label{fig:Zg_eq}
\end{figure}

We now illustrate our main result that the mass loading factor governs the
MZR below the blowout scale by applying the analytical relations derived in
\S~\ref{ssec:analytic_imp} to the vzw simulation.  In particular, we
show that our simulated galaxies' metallicities do indeed track the 
equilibrium metallicity.

For each simulated galaxy in the vzw model, we have used our progenitor 
lists to track how the ratio of its gas phase metallicity to its equilibrium 
metallicity ($X_Z \equiv Z_g/(y\sfr/\acc)$; cf.~Section~\ref{sssec:scatter}) 
varies with mass and time during the interval $6 < z < 2$.  If galaxies' 
gas-phase metallicities closely track equilibrium and if infalling material 
is unenriched then we expect $X_Z\approx1$ at all masses and redshifts.  

The result is shown in the top panel of Figure~\ref{fig:Zg_eq}.  Comparing
the least and most massive galaxy bins, we find that whereas the actual
gas metallicities of $10^{10.8}\msun$ and $10^{8.7}\msun$ galaxies
differ by $\approx0.6$ dex at $z=2$ (Figure~\ref{fig:mmr0}), their $X_Z$'s
only differ by $\approx0.2$ dex.  Moreover, the spread in $X_Z$ is even 
tighter before $z=2$.  This would not be expected in the absence of an
equilibrium condition.  The fact that the spread in these ratios is 
tighter than the spread in galaxies' actual metallicities indicates that 
at all times galaxies' metallicities are tightly constrained by a balance 
between enrichment from star formation and dilution from inflows.  
The fact that the ratios are offset from zero implies that the mean 
metallicity of inflowing gas is more than 10\% of the mean metallicity in 
the galaxy's ISM and reflects the widespread presence of galactic 
fountains in the vzw model; the ratio $\alpha_Z$ can be read 
from the y-axis on the right side of the top panel.  The increase in
$\alpha_Z$ with cosmic time reflects the growing relative contribution of
galactic fountain gas with respect to pristine ISM gas: by $z=2$, 
$\alpha_Z\approx50\%$, indicating that roughly 50\% of the infalling gas 
is galactic fountain material.  The increase in $\alpha_Z$ with mass 
reflects the fact that the rate at which pristine IGM gas accretes onto 
the galaxies declines with increasing halo mass owing to their increasing 
hot gas fractions~\citep[e.g.,][]{bir07}.

The top panel of Figure~\ref{fig:Zg_eq} also supports our view that 
pre-enrichment of gas that falls onto galaxies is not significant in 
the vzw simulation, which can be understood as follows: If, at a given 
redshift, gas that is being accreted  were homogeneously pre-enriched to 
a certain level, this pre-enrichment would provide a relatively larger 
boost to the metallicities of low-mass galaxies than to massive galaxies.  
As a result, the gas-phase metallicities of low-mass galaxies would lie
farther above their expected equilibrium value given unenriched infall, 
and the normalization of the $X_Z$ trend for low-mass galaxies would be 
boosted systematically above the $X_Z$ trend for more massive galaxies.
In fact, all but the most massive galaxies display roughly the same
ratio $X_Z$ down to $z=2$, hence a homogeneous pre-enrichment is not 
significant in this model.

In Sections~\ref{sec:metEvol} and~\ref{sec:compare_analytics} we proposed
that the rate at which gas is processed into stars and winds tracks the
rate at which it is accreted, $\sfr(1+\mlf)=\acc$.  This idea is central
to the current work as it allows us to demonstrate that the slope and 
normalization of the MZR depend almost entirely on the scaling of $\mlf$.
Additionally, a systematic imbalance between the rates of gas accretion and 
gas processing could in principle mimic the effects of nonzero $\alpha_Z$, 
leading to an incorrect interpretation of the top panel of 
Figure~\ref{fig:Zg_eq}.  For this reason, we show the ratio of the
gas processing to the gas accretion rates in our vzw model in the bottom 
panel of Figure~\ref{fig:Zg_eq}.  This figure indicates that galaxies 
generally process their gas at the same rate as they accrete it, 
justifying our assumption\footnote{Note that this is a generalization of 
the \emph{Ansatz} $\sfr = \acc$ that was introduced by~\citet{lar72}.} 
that $\sfr/\acc=1/(1+\mlf)$ and supporting our view that
the increase in $X_Z$ with time in the top panel results from galactic 
fountains rather than a mismatch between gas accretion and gas processing 
rates.  Interestingly, this plot also indicates that, in the vzw model, 
mergers do not contribute significantly to the buildup of galaxies' stellar
populations even in our most massive bin.  Because we measure baryonic
accretion rates rather than strict gas accretion rates (the latter being
difficult to infer from our simulations), a tendency for galaxies to
accrete a significant fraction of their baryons as ready-formed stars
would show up as a tendency for the accretion rate to exceed the gas
processing rate.  There is some evidence that this does occur in the
most massive galaxies in the vzw model, i.e. that dry mergers are more
prevalent at high masses.  However, on average galaxies at $z\geq2$ do 
not accrete more than $\approx10\%$ of their baryons in the form of 
already-formed stars~\citep[see also][]{gw07}.

In summary, the top panel of Figure~\ref{fig:Zg_eq} verifies that the 
slope and normalization of the observable MZR are dominated by the
equilibrium condition in Equation~\ref{eqn:Zg_eq}, while the bottom
panel verifies that galaxies process newly accreted gas into stars and
winds at roughly the gas accretion rate.  In other words, at all masses 
and redshifts, metallicities are dominated by an equilibrium between the 
rates of enrichment and dilution while the enrichment rate is dominated 
by an equilibrium between the rates of gas accretion and gas processing.  
These equilibrium rates are governed primarily by the mass loading factor,
hence the scaling of the mass loading factor directly determines the 
scaling of the MZR.

\subsection{Scatter} \label{ssec:compare_scatter}
\begin{figure}
\setlength{\epsfxsize}{0.5\textwidth}
\centerline{\epsfbox{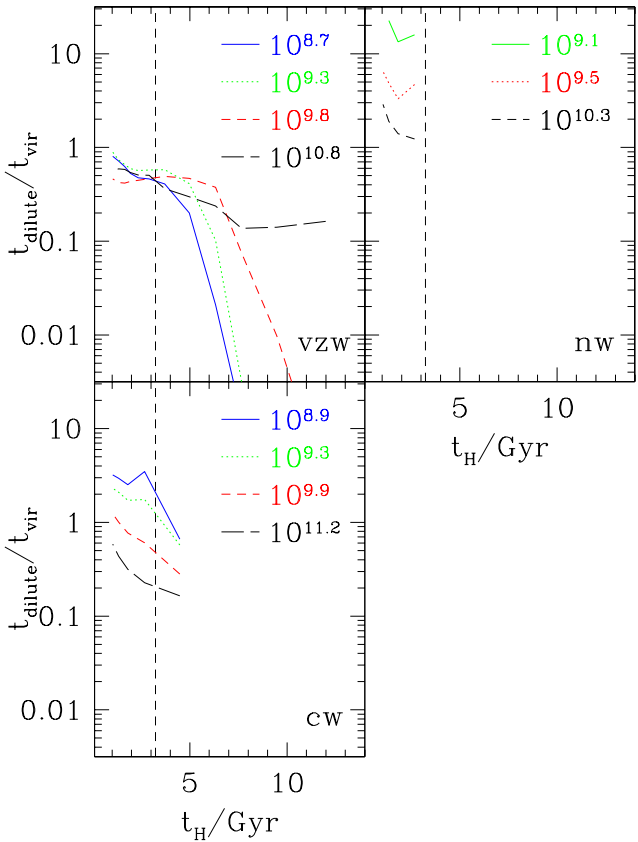}}
\vskip -0.0in
\caption{The ratio of the dilution time $M_g/\acc$ to the dynamical time as a 
function of mass and time in our various models.  The galaxy masses at $z=2$
are indicated in the legend and the vertical dashed lines indicate $z=2$.  
Comparison with Figure~\ref{fig:mmr0} indicates that mass scales with short 
dilution times tend to show small scatter in the MZR whereas mass scales 
with long dilution times tend to show large scatter.
}
\label{fig:t_dilute}
\end{figure}
In our analytical model, scatter in the MZR occurs because a perturbation,
such as an accretion or merger event, displaces a galaxy's metallicity
from its equilibrium value.  The timescale to return to equilibrium is
given by the gas dilution time $M_g/\acc$.  Perturbations to a galaxy's
metallicity are expected to occur on timescales no shorter than the
dynamical time, $t_{\rm vir} \simeq 2.5 {\rm Gyr} / (1+z)^{3/2}$.
Hence if the dilution time is shorter than the dynamical time, then we expect
perturbations to disappear rapidly and for the scatter in the MZR to
be tight.

Figure~\ref{fig:t_dilute} shows how the ratio of the dilution time to
the dynamical time varies with scale and time in our models.  In the
nw models the dilution times scale inversely with mass at early times
from roughly $1.5 t_{\rm vir}$ in the most massive galaxies to roughly
$15 t_{\rm vir}$ in the lowest-mass galaxies at $z\sim$2--3.  In the cw
model the dilution time scales more slowly with mass although the trend
for low-mass galaxies to dilute their gas reservoirs more slowly than
massive galaxies is preserved despite the cw winds.  Generally, galaxies
less massive than $10^{10}\msun$ have dilution times longer than $t_{\rm
vir}$ at $z=2$ while more massive galaxies have dilution times that are
shorter than $t_{\rm vir}$.  Finally, in the vzw model the dilution time
does not exceed $t_{\rm vir}$ at any mass scale.  More interestingly,
in this model it is the low-mass galaxies whose dilution times are the
shortest, a direct consequence of our assumption that the mass loading
factor scales with the inverse of the velocity dispersion.

Turning to the question of the scatter in the MZR (Figure~\ref{fig:mmr0}),
the relatively long dilution times in the nw model at $z=2$ are consistent
with the relatively large scatter ($\approx0.11$ dex) in the corresponding
MZR: Because perturbations to these galaxies' metallicities disappear
relatively slowly, they spend a relatively large amount of time out of
equilibrium.  Galaxies in the vzw model possess significantly shorter
dilution times than the nw galaxies or the low-mass cw galaxies,
consistent with the tight scatter ($\approx0.08$ dex) in the vzw MZR
at all scales.  In the cw model low-mass galaxies possess long dilution
times and large scatter whereas massive galaxies possess short dilution
times and small scatter, as expected.  The most massive cw galaxies
possess the shortest dilution times of any of our simulated galaxies at
$z=2$, consistent with the extremely tight scatter ($\approx 0.06$ dex)
in the cw MZR above the blowout scale.  The dilution times for low-mass
cw galaxies are shorter than they are for the low-mass nw galaxies even
though the scatter in the MZR is largest for the low-mass cw galaxies.
The extra scatter at low masses owes to these galaxies having had their
baryons stripped by winds from neighboring galaxies.  Because the amount
of stripping varies with environment, it effectively creates a range of
equilibrium metallicities for each halo mass.

In summary, galaxies tend to process gas into stars and winds at
roughly the gas accretion rate.  This tendency leads to the existence
of an equilibrium gas-phase metallicity $\zeq = y\sfr/\acc$, which
encodes information regarding both a galaxy's present conditions (via
the accretion rate and wind properties) and its star formation history
(via the current gas content, which determines $\sfr$).  Corrections to
this zeroth-order equilibrium result from galaxies accreting pre-enriched
gas or ready-formed stars as well as effects that depend on environment;
the first two of these effects should not increase the scatter in the
MZR while the last one should.  Metallicities are expected to lie close
to $\zeq$ as long as dilution times are short compared to a dynamical
time; in our momentum-driven wind model that achieves the best agreement
with the observed $z=2$ MZR, this holds for all scales and epochs.
Our interpretation of the origin of the MZR scatter also explains the
no-wind and constant wind cases.  Together with its relative simplicity,
we believe this makes our interpretation reasonably compelling.

\section{Summary} \label{sec:Summary}

In this paper we have compared the observed MZR of star-forming
galaxies at $z\sim2$ with predictions from cosmological hydrodynamic
simulations that incorporate three different models for galactic
outflows: No outflows, a ``constant wind" model that emulates the
energy-driven~\citet{dek86} scenario, and a ``momentum-driven wind" model
that reproduces $z\ga 2$ IGM metallicity observations~\citep{opp06}.
We have shown that the momentum-driven wind model produces the best
agreement with the slope, normalization, and scatter of the observed MZR.
We have constructed a simple analytical model that qualitatively reproduces
the behavior of our simulations, and used it to identify the processes
that drive galaxies' metallicities.  Our work shows that the slope,
normalization, and scatter of the MZR as well as its evolution with time
all constitute constraints on the behavior of outflows.  In particular,
our simulations strongly disfavor any constant wind scenario, and explain
why our momentum-driven wind model produces reasonable agreement with
available constraints.  Our main conclusions are summarized as follows:

\begin{itemize}
\item Outflows are required in order to bring the simulated and observed 
MZRs into agreement at $z\approx 2$.  Without outflows, enrichment proceeds too
rapidly relative to dilution with the result that galaxy metallicities 
are 2--3$\times$ higher than observed.

\item The MZR is governed by an evolving equilibrium between the
enrichment rate owing to star formation and the dilution rate owing to
gas accretion.  This results in an {\it equilibrium metallicity} for any given
galaxy, given by $\zeq = y\sfr/\acc$, where $y$ is the yield, $\sfr$
is the star formation rate, and $\acc$ is the gas accretion rate.

\item Outflows affect $\zeq$, and hence the MZR, by limiting the gas
supply for star formation.  For a given mass loading factor $\mlf$,
and assuming (as our simulations predict) infalling gas has a negligible
metallicity, then $\zeq=y/(1+\mlf)$.

\item Wind speeds ($\vw$) affect the MZR by governing how much outflowing
mass actually escapes the halo.  This results in an effective mass
loading parameter $\emlf$ (Figure~\ref{fig:eta_eff}), which is similar to (but
slightly less than) $\mlf$ so long as winds are fast enough to escape a
galaxy's halo, and drops rapidly towards zero for galaxies whose winds are
slower than the escape speed.  Our simulations' metallicities are hence
well described by $\zeq\approx y/(1+\emlf)$, where $\emlf\approx\mlf$
below the blowout scale, and $\emlf\approx 0$ above it.

\item The reheating scale, which is the mass below which the outflow
energy input is sufficient to unbind all the gas, does not play a
significant role in determining $\emlf$.  This is evident because in our
cw run the fraction of baryons converted to stars ($f_*$) does not vary
with halo mass in the way expected under the assumption of efficient
energetic coupling of outflows with ambient gas (Figure~\ref{fig:eta}).
Physically, this is because in our simulations outflows tend to blow
holes in surrounding gas rather than heat it.

\item The observed slope and amplitude of the MZR therefore constrain
how $\emlf$ and $\vw$ varies with $M_*$.  Our momentum-driven wind model
obtains the observed relation $Z_g(M_*)\propto M_*^{0.3}$ by having
$\mlf\propto 1/\sigma \propto M_{\rm halo}^{-1/3}\propto M_*^{-1/3}$,
and by having outflow speeds always above the escape velocity (so
$\emlf\approx\mlf$).  The latter constraint requires some positive mass
dependence of outflow speeds on galaxy mass, which we assumed to be
$\vw\propto\sigma$, but any dependence where galaxy masses are always
below the blowout scale would suffice.

\item Our no-wind scenario also produces a MZR with roughly the correct
slope, though in detail it is too shallow.  In the absence of outflows,
this owes to a mass dependence of $M_*$ on $M_{\rm halo}$ such that
low-mass galaxies have a lower fraction of baryons in stars.  Hence in
principle it is possible to match the observed MZR without having $\emlf$
vary with $M_*$.  However, the required scaling of $M_*$ with $M_{\rm
halo}$ does not occur naturally in our simulations with outflows.

\item Our constant wind scenario fails to even qualitatively
match the observed MZR.  The existence of a blowout scale at $\sim
10^{10}M_\odot$ produces a marked feature in the MZR, below which
$\emlf\approx\mlf=$constant, and above which $\emlf$ goes rapidly
to zero and hence the MZR rises quickly towards $\zeq=y$.  Such a
feature is generically expected across the blowout scale.  The absence
of such a feature in the observed $z\approx 0$ MZR from $M_*\approx
10^7-10^{11}M_\odot$ argues against a blowout scale in that mass range,
thereby ruling out any reasonable constant wind speed scenario.

\item The scatter in the MZR is governed primarily by the dilution time
$t_{\rm d}=M_g/\acc$ compared to the dynamical time $t_{\rm vir}$.  If
dilution times are short compared to a dynamical time, then perturbations
from $\zeq$ have time to equilibrate, thereby suppressing scatter.
The small scatter seen in the MZR argues for $t_{\rm d}/t_{\rm vir}\la 1$
across the full range of observed masses.  Our momentum-driven wind model
satisfies this non-trivial constraint, whereas our other models do not
(Figure~\ref{fig:t_dilute}).

\item Another physical effect that plays a secondary role in governing
the MZR is that outflows carry significant amounts of momentum 
that can strip baryons from neighboring halos.  This increases
scatter by causing equilibrium metallicities to depend on environment as
well as mass.  Hence the tight scatter in the observed MZR suggests that
either galactic winds do not carry significant amounts of momentum out of
galaxies, or outflows must be sufficiently mass-loaded to ``drown out"
the effects of baryonic stripping.  In our constant wind case below the
blowout scale, neither are true, and the scatter increases significantly.
In our momentum-driven wind case, the latter is generally true.

\item Outflows and inflows cause galaxies to evolve more shallowly
than closed-box models at early times, with the vzw galaxies evolving
most shallowly of all (Figure~\ref{fig:zfg}).  Effective yields are
expected to be $\sim0.01$ at $z=2$ for both our wind models.  However,
the detailed scaling of the vzw model's $y_{\rm{eff}}$ suggests better
agreement with the well-constrained low-redshift observations.  It is
worth noting that the effective yield is only reflective of the recent
history of gas and metal accretion over a dilution time, hence it cannot
be used to infer long-term accretion histories.

\end{itemize}

According to our analysis, an outflow model that will successfully
reproduce the observed MZR must satisfy three main conditions:
\begin{enumerate}
\item $\mlf\propto$(slope of MZR)$^{-1}$ when $\mlf\gg 1$;
\item $\vw$ must scale with mass such that all galaxies are below the 
blowout scale (so that $\emlf\approx\mlf$); 
\item Dilution times must be short compared to dynamical times in order 
to maintain a small MZR scatter at all masses.  
\end{enumerate}
These criteria show that the MZR mostly constrains the mass loading factor, with
weaker constraints on outflow speeds and gas accretion rates.
It is interesting that our momentum-driven wind scenario naturally 
satisfies these requirements
(along with secondary requirements such as the subdominance of baryon
stripping).  Although other wind models could conceivably be postulated
that also satisfy these requirements, it is compelling that this same
model also satisfies IGM metallicity constraints, and broadly agrees
with available direct measurements of outflow parameters at high and low
redshift.  In any case, other wind models will likely need to satisfy
the above criteria in order to match the observed MZR, demonstrating
that the MZR provides strong constraints on outflow properties.

The rather dramatic failure of our constant wind scenario is suprising
in light of the apparent success of the simple analytical models presented
by~\citet{dek86} and~\citet{dek03}.  The root difference traces back
to those works assuming that feedback suppresses star formation by
efficiently coupling supernova energy with baryons in halos, while
our three-dimensional simulations produce inefficient coupling with a
propensity for winds to blow holes in surrounding gas.  This is partly a
result of the way we implement winds in our simulations by turning off
hydro forces for some distance; however, in practice that distance is
much smaller than the halo size and hence interactions with halo gas can
(and do) still occur.  Regardless, the existence of a strong feature
at the blowout scale seems an unavoidable consequence in a constant wind
scenario, and is in direct conflict with the observed unbroken MZR power
law over four orders of magnitude in stellar mass.  Furthermore, the
increased scatter below the blowout scale predicted by such a scenario
is not seen.  Hence we strongly disfavor this explanation of the MZR.

The slow turnover in the $z\sim 0$ MZR at $M_*\ga 10^{10.5}M_\odot$
cannot be studied directly in our simulations owing to a lack of
sufficient dynamic range, along with the fact that our simulations were
not evolved to $z=0$.  However, it arises naturally in our scenario when
the mass loading factor becomes $\ll 1$, which yields $\zeq\rightarrow
y=$constant.  In principle, it could also arise if galaxies with
$M_*\ga 10^{10.5}M_\odot$ are above the blowout scale (which would
also make $\emlf\rightarrow 0$); indeed, this is the conventional
interpretation~\citep[e.g.][]{tre04}.  However, this would imply the
existence of a blowout scale at that mass, which as we have argued
above causes other features in the MZR that contradict observations.
Hence we suggest that this mass scale does not reflect a characteristic
wind speed, but rather a characteristic scale of the mass loading factor,
namely the galaxy mass where the mass loading factor is roughly unity.

Our findings agree broadly with those from the higher-resolution study
of~\citet{bro06}, though there are some differences in interpretation.
In their work, they determined that winds affect the MZR of low-mass
galaxies in the following sense: When they compared gas-phase
metallicities at $z=0$ with the mean metallicity of all gas that had
ever belonged to the galaxies, they found no systematic offset.  This is
expected if the mean metallicity in an outflow equals the mean metallicity
in the galaxy's ISM.  On the other hand, by comparing simulations with
and without winds they found that winds suppress gas densities and hence
star formation efficiencies, which in turn shapes the observable MZR.
Hence they deduced that star formation efficiency is a key driver of
the MZR.  We also find that more massive galaxies have more efficient
star formation in Figure~\ref{fig:simfits}, and Figure~\ref{fig:nw_comp_sfr}
shows that this is important for establishing the MZR, at least in
the no-wind case.

However, in our wind models we find that the star formation efficiency 
doesn't by itself determine the MZR, because the trends in 
Figure~\ref{fig:eta} don't mimic those of the MZR.  Instead, the MZR's 
trend is the mass scaling of $\zeq$, which is set by how the accretion rate 
compares with the star formation rate (\citealt{tl78}; eqn.~\ref{eqn:Zg_eq}); 
in our models, this is similar across all masses and close to unity at 
$z\sim 2$, when the dependence on $\mlf$ is taken into account, as shown 
in the bottom panel of Figure~\ref{fig:Zg_eq}.  Hence the mass dependence 
in $\zeq$ arises mainly from the mass dependence in $\emlf$.  In this way,
the trends in $\emlf$ (Figure~\ref{fig:eta_eff}) are directly reflected 
in the MZR.

This interpretation can be compared with the results of~\citet{kob07}, who
observed a tight correlation between stellar metallicity and the mass 
fraction of metals retained by the galaxies in their models at all 
redshifts (their Figure 16d) and concluded that higher stellar 
metallicities result directly from a lower mass fraction of metals 
ejected.  Our models obey a similar correlation, hence our conclusions 
should be consistent with theirs.  In fact, the tendency of gas-phase 
metallicities to track an equilibrium value combined with a tendency 
for $\emlf$ to decline and $\zeq$ to grow as galaxies grow~\emph{requires}
that the fraction of metals retained and the mean stellar metallicity 
must grow together, as found by~\citet{kob07}.  Moreover, given that the 
gas-phase MZR shows little scatter at all redshifts, the stellar MZR should 
also show little scatter (as also noted by~\citealt{kob07}) even though it 
is not directly governed by an equilibrium condition analogous to 
Equation~\ref{eqn:Zg_eq}.  In this way, Equation~\ref{eqn:Zg_eq} likely 
governs the gas-phase MZR directly and the stellar MZR indirectly in both 
sets of models.  Indeed, it is likely that an analogous equilibrium 
condition governs the MZR of \emph{any} galaxy evolution model that 
incorporates a treatment for strong ($\emlf\gg1$) outflows.

Despite considerable progress over the last decade, the use of
metallicities and effective yields to constrain galaxy evolution
constitutes a field that is in its infancy from both theoretical and
observational perspectives.  Our simulations' implementation of outflows,
while being state-of-the-art for cosmological simulations, is still
crude.  For instance, we currently assume enrichment only from Type~II
supernovae, we do not shut off winds in galaxies with low star formation
rate surface densities~\citep{hec03}, and we use the local potential
as a proxy for galaxy mass.  All of these simplifications are probably
not fatal at $z=2$, but by $z=0$ they likely are; this (in addition to
computer time constraints) is the main reason we have not attempted
to extend our simulation analysis to $z=0$.  We are working towards
incorporating metals from Type~Ia supernovae and stellar mass loss,
improved wind criteria, and direct galaxy identification into our code
(Oppenheimer et al., in preparation).  Our preliminary results indicate
that none of these issues significantly impact the predicted $z=2$ MZR.

Another aspect for future exploration is different scalings of the wind
model.  For instance, our constant wind scenario is only one possible
implementation of energy-driven outflows.  More sophisticated versions
that allow $\mlf$ and the wind speed to vary could improve the agreement
between the observed and simulated luminosity functions at the faint end
while yielding agreement with the observed MZR.  It is by no means clear
that momentum-driven winds, as we have implemented them, are the only
viable alternative.  Indeed, it is for this reason intriguing 
that~\citet{kob07} have obtained reasonable agreement with the observed 
MZR using a treatment for \emph{pressure}-driven outflows that result in 
a qualitatively similar scaling of mass-loading factor versus mass as
well as the predicted MZR.  The intuition gained from these results can 
hopefully guide us (and others) towards understanding how alternative 
outflow models may fare prior to running expensive simulations.

On the observational side, galactic outflows are still relatively
poorly constrained despite impressive advances over the past decade.
More detailed measurements of the mass loading factors and wind
speeds across a large dynamic range would be helpful in indicating
whether momentum-driven or energy-driven winds are likely to dominate.
More importantly, despite heroic observational efforts there remain
relatively few constraints on galaxies' metallicities and gas fractions at
high redshift.  Upcoming metallicity measurements made with multi-object
infrared spectrographs such as FLAMINGOS-2 as well as direct gas mass
measurements made with IRAM and ALMA will prove crucial in finally
allowing us to apply these metrics to the high-redshift Universe.

Despite its simplicity, the fact that our model explains the detailed
shapes of both of our wind models' MZRs leads us to believe that
it captures most of the essential physics.  Our scenario invokes two
parameters, the equilibrium metallicity and the dilution time, neither of
which can be directly measured because they depend on the gas accretion
rate.  Instead, they must be constrained indirectly through observations
of how galaxy properties such as SFR, gas mass, and metallicity vary
with mass and epoch.  We look forward to undertaking such comparisons to
observations in the future, and are hopeful that they will shed further
light on the critical problem of understanding galactic outflows.

\section*{Acknowledgements}

We thank V. Springel and L. Hernquist for providing us with Gadget-2.  
We thank Christy Tremonti, Don Garnett, John Moustakas, Alison Coil, 
Nicolas Bouch{\'e}, Ben Oppenheimer, and Daniel Eisenstein for assistance 
and helpful discussions.  We thank the anonymous referee for many helpful 
comments that improved the paper.  Partial support for this work was provided
by NASA through grant numbers HST-AR-10647 and HST-AR-10946 from the 
Space Telescope Science Institute, which is operated by the Association 
of Universities for Research in Astronomy, Inc., under NASA contract 
NAS5-26555.  Partial support for this work, part of the Spitzer Space 
Telescope Theoretical Research Program, was provided by NASA through 
a contract issued by the Jet Propulsion Laboratory, California 
Institute of Technology under a contract with NASA.  Our simulations 
were run on the Xeon Linux Supercluster at the National Center for 
Supercomputing Applications.  KMF acknowledges support from a National 
Science Foundation Graduate Research Fellowship.

\twocolumn

\end{document}